\documentclass{article}

\usepackage{arxiv}

\usepackage[utf8]{inputenc} 
\usepackage[T1]{fontenc}    
\usepackage{amsthm}
\usepackage{amsmath}
\usepackage{amssymb}
\usepackage{natbib}
\usepackage[colorlinks,citecolor=blue,urlcolor=blue,filecolor=blue,backref=page]{hyperref}
\usepackage{graphicx}
\usepackage{natbib}

\usepackage[dvipsnames]{xcolor}
\usepackage{dsfont} 
\usepackage{enumerate}
\usepackage{caption}
\numberwithin{equation}{section}
\theoremstyle{plain}

\newcommand{\E}{\mathbb{E}}
\renewcommand{\P}{\mathbb{P}}
\newcommand{\Q}{\mathbb{Q}}
\newcommand{\Var}{\mathbb{V}}
\newcommand{\Cov}{\mathbb{C}\mathrm{ov}}

\newcommand{\btheta}{\boldsymbol{\theta}}

\title{Scalable Approximate Bayesian Computation for Growing Network Models via Extrapolated and Sampled Summaries}

\author{
    Louis Raynal\\
    Department of Biostatistics\\
    T.H. Chan School of Public Health, Harvard University\\
    655 Huntington Avenue, Building 2, 4th Floor\\
    Boston, MA, USA 02115\\
    \texttt{llcraynal@hsph.harvard.edu}\\
    \And
    Sixing Chen \thanks{Shared first authorship}\\
    Department of Biostatistics\\
    T.H. Chan School of Public Health, Harvard University\\
    655 Huntington Avenue, Building 2, 4th Floor\\
    Boston, MA, USA 02115\\
    \And
    Antonietta Mira\\
    Data Science Lab, Institute of Computational Science\\
    Universit\`a della Svizzera italiana\\
    Via Buffi 6, 6900 Lugano, Switzerland\\
    Dipartimento di Scienza e Alta Tecnologia\\
    Universit\`a degli Studi dell'Insubria\\
    Via Valleggio, 11 - 22100 Como, Italy\\
    \And
    Jukka-Pekka Onnela\\
    Department of Biostatistics\\
    T.H. Chan School of Public Health, Harvard University\\
    655 Huntington Avenue, Building 2, 4th Floor\\
    Boston, MA, USA 02115\\
}

\begin{document}
\maketitle

\begin{abstract}
Approximate Bayesian computation (ABC) is a simulation-based likeli\-hood-free method applicable to both model selection and parameter estimation. ABC parameter estimation requires the ability to forward simulate datasets from a candidate model, but because the sizes of the observed and simulated datasets usually need to match, this can be computationally expensive.
Additionally, since ABC inference is based on comparisons of summary statistics computed on the observed and simulated data, using computationally expensive summary statistics can lead to further losses in efficiency. ABC has recently been applied to the family of mechanistic network models, an area that has traditionally lacked tools for inference and model choice. Mechanistic models of network growth repeatedly add nodes to a network until it reaches the size of the observed network, which may be of the order of millions of nodes. With ABC, this process can quickly become computationally prohibitive due to the resource intensive nature of network simulations and evaluation of summary statistics.
We propose two methodological developments to enable the use of ABC for inference in models for large growing networks. First, to save time needed for forward simulating model realizations, we propose a procedure to extrapolate (via both least squares and Gaussian processes) summary statistics from small to large networks. Second, to reduce computation time for evaluating summary statistics, we use sample-based rather than census-based summary statistics. We show that the ABC posterior obtained through this approach, which adds two additional layers of approximation to the standard ABC, is similar to a classic ABC posterior. Although we deal with growing network models, both extrapolated summaries and sampled summaries are expected to be relevant in other ABC settings where the data are generated incrementally.
\end{abstract}


\section{Introduction}

Networks are used to study systems where individual agents or elements do not operate in isolation but instead have complicated structural or functional connections with other elements in the system. There are currently at least two paradigms to model network structure. \emph{Statistical network models} directly model the observed network data. Their likelihood functions are usually available in closed form (possibly up to a normalizing constant), and inference and model selection tools are generally available. Perhaps the best known example of this model class is the family of exponential random graph models (ERGM) \citep{lusher2013exponential}. In contrast, \emph{mechanistic network models} are algorithmic descriptions of network formation, and they are defined by a small number of domain-specific rules that are informed by our scientific understanding of the problem. Their likelihood functions are generally not analytically tractable, and thus inference and model selection tools have traditionally not been developed for them. In network science, the origins of which are primarily in physics, there are easily hundreds of models like this. In fact, for a long time mechanistic models were essentially the only type of models that were formulated and studied, using both mathematical methods and computer simulation. Well-known examples of this model class include the Price model \citep{price1965networks}, the Barab\'{a}si-Albert model \citep{barabasi1999emergence}, the Watts-Strogatz model \citep{watts1998collective}, and many others \citep{sole2002model,vazquez2003modeling,klemm2002highly,kumpula2007emergence}.

In many settings, mechanistic models, when contrasted with statistical models of network data, can better address questions of scientific interest as they allow the inclusion of a small number of known mechanisms. Performing model selection and parameter inference on mechanistic network models can therefore help us select among competing sets of hypotheses (sets of mechanisms) and assess how adequate the given mechanisms are for explaining the observed networks. Mechanistic models may of course be wrong just as our scientific understanding of a phenomenon may be wrong, but mechanistic models allow one to test the merit of different hypotheses with the goal of discarding those that are not in agreement with data. For example, one hypothesis might be to assess whether some form of preferential attachment, a commonly postulated mechanism of network growth, is needed explain the structure of the observed network \citep{barabasi1999emergence}. The engagement of mechanistic models with real-world data has traditionally been shallow due to a lack of statistically sound inferential and model selection tools. To address some of the gaps in methodology for mechanistic models, we previously introduced a general approximate Bayesian computation-based framework for inference and model selection \citep{onnela2018statistical}, a flexible model selection framework for mechanistic network models \citep{chen2018flexible},
a bootstrap method for goodness of fit and model selection with a single observed network \citep{chen2019bootstrap}, a framework for converting mechanistic network models to probabilistic models \citep{goyal2020framework}, a Bayesian inference scheme for spreading processes on networks \citep{dutta2018bayesian}, and have used standard machine learning methods for feature-based classification of networks \citep{barnett2016feature}. To this end, we have developed a user-friendly, extensible, and parallel library for ABC in Python \citep{dutta2017abcpy}.

Approximate Bayesian computation (ABC) is a simulation-based and likelihood-free method with wide applicability to settings with intractable likelihoods for both parameter estimation and model selection \citep{Marin2012,sunnaaker2013approximate,lintusaari2016fundamentals}. Many of the approaches mentioned above make use of ABC since mechanistic network models typically have intractable likelihood functions. The application of ABC for parameter estimation requires the ability to forward simulate synthetic datasets from candidate models. A key feature of ABC is that inference is based on comparisons of summary statistics computed on the observed and simulated data, and therefore efficient evaluation of summary statistics is another important practical requirement for ABC. A characteristic of many mechanistic network models is that they grow the network starting from a small seed network, often one node at a time, until a predetermined network size (number of nodes) is reached, at which point the process terminates. These types of models, a subset of all mechanistic network models, are often called \textit{growing network models}. Because many of the studied networks are large, from thousands to millions of nodes, the use of ABC for growing network models introduces two potential computational bottlenecks: the cost of simulating networks and the cost of computing summary statistics. Using $n$ to denote the number of nodes in the network, let $O(n^\alpha)$ denote the computational complexity of network simulation and $O(n^\beta)$ denote the complexity of summary statistic computation. Network simulation is relatively fast for most mechanistic models, and it appears that typically $\alpha \le 2$. Summary statistic computation, in contrast, is greatly dependent on the specific summary. For example, it is possible to evaluate the so-called betweenness centrality, a global measure of network connectivity, in $O(n(m+n))$, where $m$ denotes the number of edges in the network \citep{newman2010networks}. A more complex example is that of triangle enumeration, which is trivially solvable in $O(n^3)$, whereas the best-known algorithm takes time $O(n^{2.373})$ on sparse power-law graphs using fast matrix multiplication \citep{latapy2008main}.
Even more complex summaries, such as identification of network community structure \citep{brandes2006maximizing,fortunato2010community,traag2011narrow}, can be NP-hard problems, in practice requiring the use of heuristics for their evaluation.

Methods to alleviate the computational bottleneck from forward-simulation in ABC have been previously explored outside of the network setting. Methods by \citet{gutmann2016bayesian} and \citet{pmlr-v33-wilkinson14} approximate the log-likelihood
function of the summary statistics, while \citet{moores2015pre} uses a binding function to create a map between parameters of the
intractable likelihood function of the summary statistics and parameters of a tractable likelihood function of an alternative
model in order to approximate the former likelihood with the latter. These approaches try to save computation by using
forward-simulation (from a limited set of parameter samples from the prior) to build an approximation to the likelihood function,
with which one can compute the posterior distribution from a larger parameter sample without the need for further forward-simulation.
Methods by \citet{conti2010bayesian} and \citet{carbajal2017appraisal} directly approximate the summary statistics output
of a complex simulator with a computationally efficient Gaussian process-based emulator.
In either case, these previous methods connect the summary statistics to the parameter values at a fixed size of the dataset, but do not consider summary statistics for growing datasets. The notion of a growing dataset is an important albeit not unique feature of the network setting, and forms the basis of our approach.

We propose two methodological developments to make ABC feasible for modeling large empirical networks with growing mechanistic models. First, to save time needed to simulate model realizations, we propose a procedure to extrapolate summary statistics from small to large networks. Rather than growing the network to $n_o$ nodes as in the observed graph, we propose to stop at some $n_s \ll n_o$ and extrapolate the summary from $n_s$ to $n_o$. Second, to save time needed evaluating summary statistics, we propose using sample-based rather than census-based summary statistics. For example, rather than counting all triangles in a network, we count the number of triangles within a subset of $n^*$ nodes, where  $n^* \ll n_o$. To illustrate the implications of the former, consider a situation where $\alpha = 2$; if we can stop simulation early at $n_s = n_o / 10$, the time required to simulate a small network of $n_s$ nodes is only 1\% of the time required to simulate one large network of $n_o$ nodes. To depict the latter, in the best case scenario it might be possible to fix the size of the subsample; for example, if one were to use average degree as a summary, full enumeration would scale as $O(n)$ whereas computing the degree only for a subsample of fixed size (say, 10,000 nodes) leads to $O(1)$ complexity. These considerations suggest that the combination of summary statistic extrapolation and use of sample-based summaries could result in significant computational savings. We note that although our problem deals with mechanistic models of growing networks, both extrapolated summaries and sampled summaries are expected to be applicable in other ABC settings as well. Moreover, our approach can be used to study any mechanistic network model; these methods are not limited to the models presented in this paper, as the only requirement is our ability to forward-simulate networks given some parameter values.

This paper is organized as follows. We introduce our method in Section \ref{methods}, discuss our results for summary statistic extrapolation in Section \ref{simulation}, and those for sample-based summary statistics in Section \ref{subsampling}. We demonstrate our method with an empirical dataset, a citation network in the physical sciences, in Section \ref{empirical}. We conclude our investigation with discussions in Section \ref{end}.

\section{Materials and Methods}
\label{methods}

\subsection{Notation}
Given an empirical network $G_{o}$ of $n_o$ nodes, our goal is inference on the parameters $\btheta$,
with prior distribution $\pi_{\btheta}$, which index the mechanistic network model $M = M(\btheta)$.
Network realizations $\tilde{G}_1, \ldots, \tilde{G}_B$,
each associated with a set of parameters drawn from $\pi_{\btheta}$, of $n_s\ll n_o$ nodes are generated from $M$ for an ABC procedure that will be based on the vector of summary statistics $\boldsymbol{S}$.
As each generated network realization $\tilde{G}_b$ grows from the seed network towards a network of $n_s$ nodes, $\boldsymbol{S}$ is evaluated at various number of nodes $n_1< \ldots < n_t < \ldots < n_s$.
Note that $\{ n_t \}$ need not be evenly spaced. $\boldsymbol{S}$ computed for the $b$th realization at $n_t$ nodes is denoted $\tilde{\boldsymbol{S}}_{b}(n_t)$ for $b=1,\ldots, B$, while that computed for the observed network $G_{o}$ is denoted $\boldsymbol{S}_{o}$. The goal is to extrapolate the value of $\boldsymbol{S}$ at $n_o$ nodes
for each $\tilde{G}_b$ based on the $\tilde{\boldsymbol{S}}_{b}(n_t)$ values, for all $n_t$. The ABC procedure is based on the reference table populated by the extrapolated quantities $\{ \hat{\boldsymbol{S}}_b(n_o) \}$. A single element of
$\boldsymbol{S}$ is denoted $s$, with corresponding notations $\tilde{s}_{b}(n_t)$ for $\tilde{G}_b$
and $s_{o}$ for $G_{o}$.

\subsection{Evolution of Network Statistics}
We consider the classic duplication-mutation-complementation (DMC) model of \citet{vazquez2003modeling}, which is used for studying protein-protein interaction networks. Note however that the method presented here can be applied to any model of network growth as the only requirement is the ability to simulate networks given some parameter values. The DMC model captures the effect of gene duplication, which is one of the primary forces behind the evolution of genomes and one of the dynamical mechanisms of network growth, although it is not the only mechanism. Gene duplication is a random event at the molecular level and can occur, for example, due to an error in DNA replication or during homologous recombination. The duplicated version of a gene is usually under less selective pressure than its parent and is therefore free to mutate rapidly, and could potentially take on a novel function. Since genes encode proteins, this new function of the gene would imply that the corresponding protein in the protein-protein interaction might acquire the ability to interact with proteins that its parent does not interact with. Put simply, gene duplication causes changes in the structure of the associated protein-protein interaction network.

The DMC model grows a network from a small seed network according to its generative mechanisms until the required number of nodes is reached. At the beginning of each step of the network generation process, an existing node is chosen uniformly at random for duplication. Edges are then added between the new node and the neighbors of the original node. Then, for each neighbor of the original node, either the edge between the original node and the neighbor or the edge between the new node and the neighbor is removed with probability $q_m$. Finally, with probability $q_c$, an edge is added between the new node and the original node. This process is repeated until the network has been grown to the desired size.

\begin{figure}[ht]
\centering
\includegraphics[ width=0.8\linewidth, trim={ 1.25cm 1.5cm 2cm 2cm }, clip ]{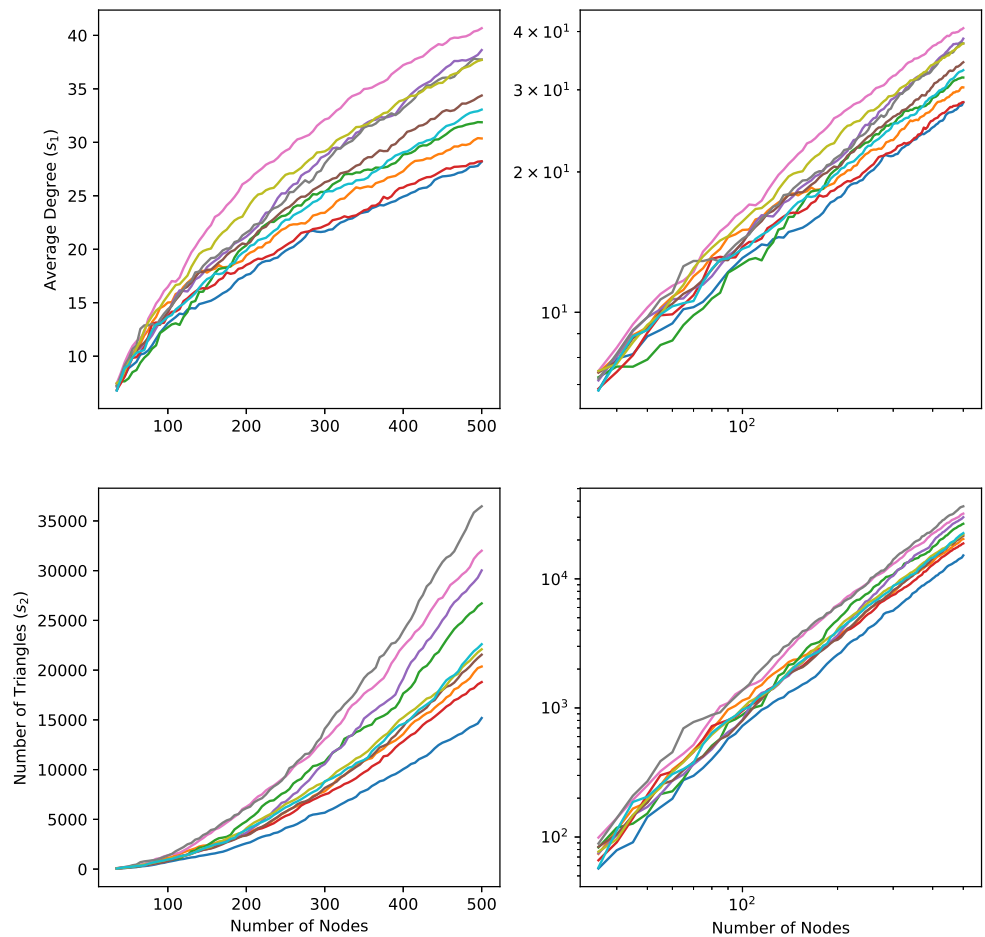}
\caption{Average degree (top) and number of triangles (bottom) computed for networks generated from the DMC model. Different colors correspond to different network realizations; linear scale on the left, and log-log on the right. An initial seed network of $30$ nodes is used to start the growth process.}
\label{fig:DMC:summary_sim}
\end{figure}

As a proof of concept, we consider two network statistics, the average degree (denoted $s_1$) and the number of triangles (the number of triplets of nodes that are all connected to one another, denoted $s_2$), for networks generated from the DMC model. Figure \ref{fig:DMC:summary_sim} shows the two statistics tracked for network realizations simulated from the DMC model.
Both statistics display non-linear growth in the number of nodes.

\subsection{Modes of Extrapolation}
The observed polynomial growth for the two investigated network statistics motivates the functional forms used in extrapolation. We assess two modes of extrapolation: a least squares fit of a polynomial function and a Gaussian process \citep{rasmussen:williams:2006} with a polynomial mean function. The former is a more barebones approach that ignores variability around some ``true'' mean as well as the correlation of the statistics at various network sizes, while the latter attempts to model both through the covariance function of the Gaussian process as described in more detail below.

\subsubsection{Least Squares Fit}
For each network realization $\tilde{G}_b$, we fit, via least squares, a function of the form $\tilde{s}_{b}(n)=a_b n^{c_b}$,
where $n$ is the number of nodes and $a_b$ and $c_b$ are realization-specific parameters to be estimated,
separately for each of the two statistics based on their corresponding tracked quantities $\{ \tilde{\boldsymbol{S}}_{b}(n_t) \}$ at $\{ n_t \}$. The functional form is based on the polynomial growth observations from Figure \ref{fig:DMC:summary_sim}, and
does not possess an intercept term since both statistics are zero at zero nodes.
We elected to fit the polynomial function on the linear scale rather than a linear function on the log-log scale, since the latter treats absolute error equally for all values of $n$. This is important since we seek to extrapolate the statistic for even larger $n_o$, and wish to minimize the absolute error at the larger network sizes.

\begin{figure}
\centering
\includegraphics[width=0.8\linewidth, trim={1.5cm 1.5cm 1.5cm 1.5cm},clip]{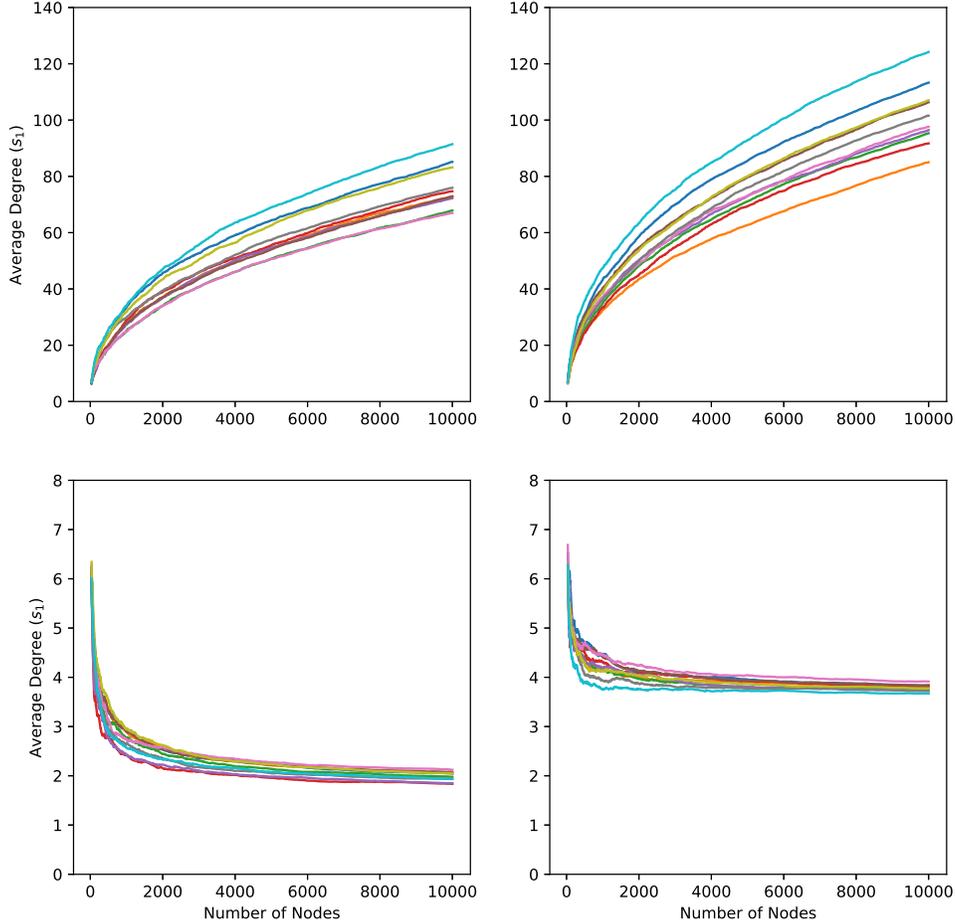}
\caption{Average degree for network realizations generated from DMC models with parameter
$(q_{m},q_{c})\in\{$(0.3, 0.3), (0.3, 0.7), (0.7, 0.3), (0.7, 0.7)$\}$
from left to right and top to bottom. Different colors correspond to different network realizations. An initial seed network of $30$ nodes is used to start the growth process.}
\label{fig:DMC:summary_sim_2}
\end{figure}

The motivation for fitting separate functions for each network realization, which do not depend on their corresponding sampled value of $\boldsymbol{\theta}$, is the large amount of variability in the networks generated by a mechanistic model, even for a given set of parameter values.  Figure \ref{fig:DMC:summary_sim_2} shows the growth of average degree for network realizations tracked for different values of $n$ for various DMC models ($q_{m}\in\{0.3,0.7\}$, $q_{c}\in\{0.3,0.7\}$).
Note the large amount of variability for even a single set of parameter values within each panel. If the function depends on the parameter and is fit with all the generated
network realizations, the extrapolated values would no longer correspond to the network realizations themselves but instead to a ``mean" value of network realizations simulated with particular parameter values. However, this variability is an intrinsic part of the model and its network realizations. Should we have generated network realizations fully to $n_o$ nodes to populate the reference table, those realizations would also contain this variability. Additionally, should the observed network truly be generated from the proposed model, then it would also contain this variability. Thus, we elected to use separate functions for each network realization that do not depend on the model parameter. Such functions follow the path the realization takes as it grows, thus reflecting the per-realization variability in both the fitted function and extrapolated value.

After fitting both functions, we evaluated them at $n_o$ to produce $\{ \hat{\boldsymbol{S}}_b(n_o) \}$,
which was entered into the ABC reference table for our procedure. Then, a standardized Euclidean distance was computed between $\boldsymbol{S}_{o}$ and each $\hat{\boldsymbol{S}}_b(n_o)$. The simulated parameter values providing the lowest distances were retained to form the samples from the ABC posterior distributions. For the remainder of this paper, LS-ABC is used to denote results using this least squares extrapolation approach.

\subsubsection{Gaussian Process} 

A Gaussian process \citep{rasmussen:williams:2006} is a stochastic process $X$ indexed by some set $T$ (typically some subset of the real line) where the finite dimensional distribution (FDD), the joint distribution of $X$ at every finite subset of $T$, is multivariate normal. A Gaussian process is defined by its mean and covariance functions $\mu$ and $K$, where for
$t_1,t_2\in T$, $\E(X(t_1))=\mu(t_1)$ and $\Cov(X(t_1),X(t_2))=K(t_1,t_2)$, where $K$ is called a kernel. We will denote such a stochastic process $GP(\mu,K)$.

In the context of extrapolating network statistics, the network statistics of $\tilde{G}_b$ are indexed by the number of nodes in the network at $\{ n_t \}$. For a given summary statistic $s$, we model $\{ \tilde{s}_{b}(n_t) \}$ as the FDD of some
Gaussian process $GP(\mu_{b}^{s},K_{b}^{s})$ at $\{ n_t \}$. The superscript and subscript of the mean and covariance
functions denote the use of separate Gaussian processes for each network statistic of each network realization.
This reflects our decision in the previous section to extrapolate each network statistic for each network realization separately. 

The mean functions of both network statistics use the same polynomial form $\tilde{s}_{b}(n)=a_b n^{c_b}$ as discussed above. On the other hand, the covariance function allows us to encode the variability of the statistics at any given $n$ as well as the covariance between the values of the statistics at any two different values of $n$. We motivate the covariance function for the two network statistics empirically.

\begin{figure}
\centering
\includegraphics[width=\linewidth, trim={3cm 0.5cm 4cm 0.5cm}, clip]{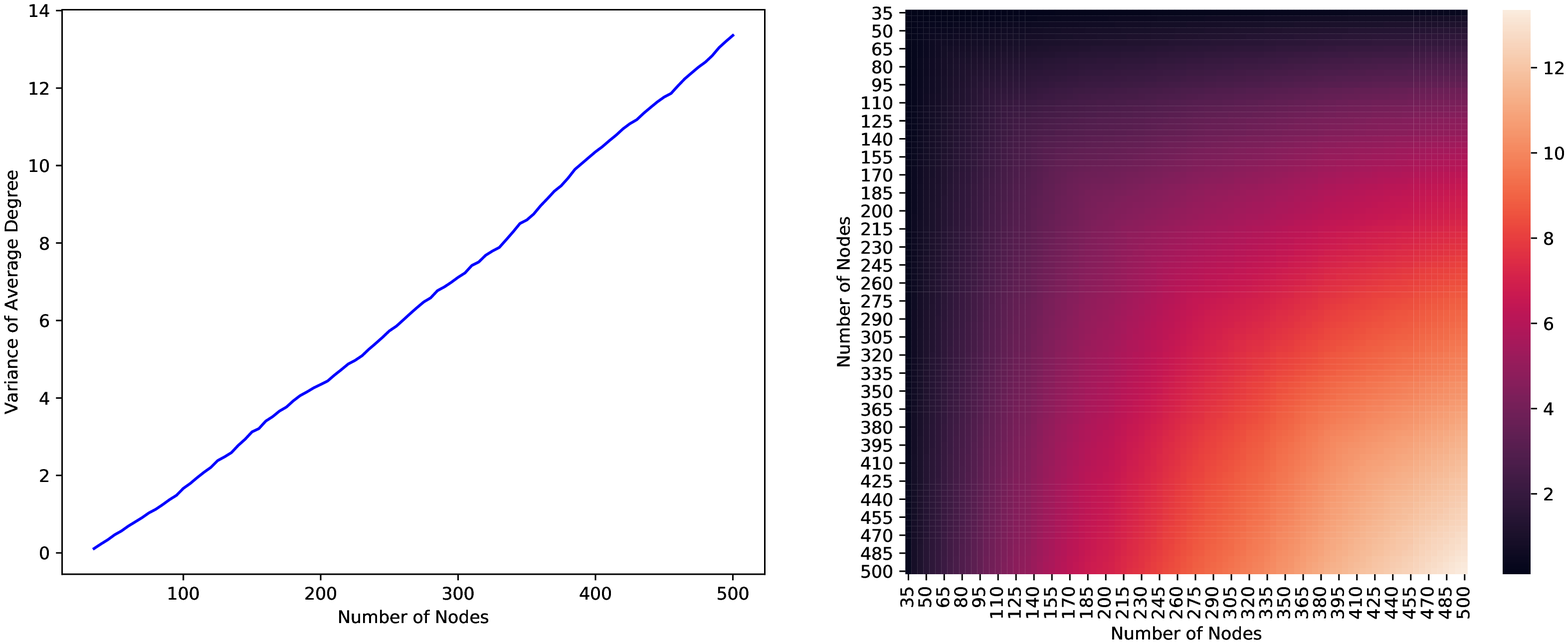}
\includegraphics[width=\linewidth, trim={3cm 0.5cm 4cm 0.5cm}, clip]{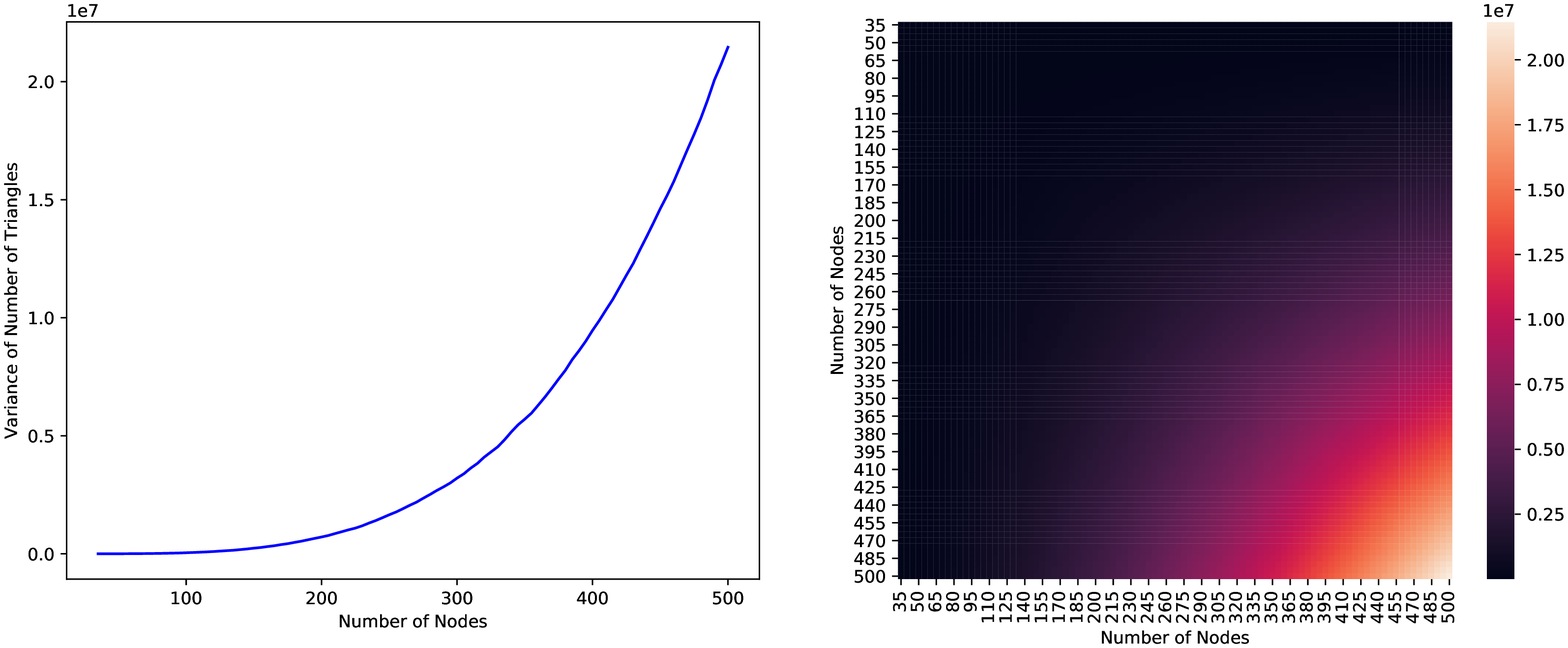}
\caption{The variance (left column) and covariance matrix (right column) of the average degree (top) and the number of triangles (bottom), for various values of $n$ over 500 network realizations generated from the DMC model with $q_m=0.5$ and $q_c=0.25$.}
\label{fig:DMC:empirical_var_covar}
\end{figure}

To investigate the covariance structure, we generated $500$ network realizations with a single set of parameters ($q_m=0.5$, $q_c=0.25$), with the two summaries tracked every 5 nodes from $35$ to $500$. The starting node number is $35$ because the seed graph for this example has $30$ nodes, as described below. 
In Figure \ref{fig:DMC:empirical_var_covar}, the left panels display the empirical variances of each summary, average degree (top) and number of triangles (bottom), and the right panels show the heatmaps of the average covariances at each pair of values of $n$ over the 500 realizations.

For both summaries, we observe in Figure \ref{fig:DMC:empirical_var_covar} that the variance increases with the number of nodes. While the growth is about linear for the average degree, it is polynomial for the number of triangles. The right panels indicate that the covariance at $n_1$ and $n_2$ grows as $n_1$ and/or $n_2$ increase. In fact, this behavior is characteristic of the linear kernel \citep[a.k.a. dot product,][p.89]{rasmussen:williams:2006} that can simply be expressed as $K(n_1, n_2) = n_1 n_2$. Such a kernel would give the desired shape for the prior variance of the triangle count, however it would not grow linearly for the average degree as required. To remedy this, we pre-multiply and post-multiply the dot product kernel by the deterministic functions $f(z)=z^{-1/2}$, with $z=n_1$ and $z=n_2$ respectively, to obtain the kernel $K'(n_1, n_2) = \sqrt{n_1} \sqrt{n_2}$. This gives a variance function increasing linearly when $n_1=n_2$.

In our experiments, to allow more flexibility to the Gaussian processes, we consider the following covariance functions for the average degree ($s_1$) and triangle count ($s_2$), evaluated at $n_1$ and $n_2$:
\begin{align*}
    \Cov(s_1(n_1), s_1(n_2)) &= (\alpha \sqrt{n_1} \sqrt{n_2} + \gamma ) + \beta \exp\left( -\frac{(n_1 - n_2)^2}{2\rho^2} \right) + \sigma^2 \mathds{1}_{\{n_1 = n_2\}},\\
    \Cov(s_2(n_1), s_2(n_2)) &= (\alpha n_1 n_2 + \gamma) + \beta \exp \left( -\frac{(n_1 - n_2)^2}{2\rho^2} \right) + \sigma^2 \mathds{1}_{\{n_1 = n_2\}},
\end{align*}
with $\alpha$, $\gamma$, $\beta$, $\rho$, $\sigma^2$ some positive parameters, specific to each kernel, and $\mathds{1}$ being the indicator function. 
The left-most term represents a parameterized version of the linear kernel, while the middle term is the well known radial basis function kernel (a.k.a. squared exponential), and the right-most term corresponds to noise on the variance terms. Results from additional kernels are presented in Appendix \ref{appendix:subsec:A.1}, but they showed no significant improvement over those presented here.

In order to estimate the parameters of the mean and covariance functions for a given network realization, we fit a Gaussian processes for each statistics separately through STAN \citep{carpenter2017stan} by specifying their corresponding FDD at $\{ n_t \}$ and a prior for each parameter.
The priors for the parameters of the mean function are normal distributions centered at their corresponding least squares fit estimates for both statistics.
For the covariance function, we assume a positively truncated standard normal distribution for the prior.
This makes sense for $\sigma^2$ as it is a noise term. 
Moreover, since $\gamma$ plays the role of an intercept term, we expect it to be approximately zero given the shape of the empirical variances (Figure \ref{fig:DMC:empirical_var_covar}, left panels). 
For the other parameters, their priors were chosen to guarantee convergence of the No-U-Turn sampler performed by STAN.
For $\alpha$ we additionally imposed a minimal value of $0.05$ to facilitate convergence of the STAN sampler, and this minimal value also guarantees that the covariance function contains a non-vanishing dot product term that is essential to mimic the empirical variance-covariance observed in Figure  \ref{fig:DMC:empirical_var_covar}.
For $\rho$, since it corresponds to the typical distance between turning points in the summary statistic curves, we constrained it to be no less than the distance between two consecutive grid points for the number of tracked nodes.

Once estimates are obtained from STAN, we evaluate the mean and covariance functions at $n_o$ to deduce the distribution of each statistic for each network realization at $n_o$. 
Note that we have two marginal normal distributions, one per summary statistic.
We can employ these distributions in different ways to enrich the reference table:
\begin{enumerate}[(i)]
    \itemsep0em 
    \item for each GP (one per summary), evaluated at $n_o$, we use the posterior mean of the GP that we take as our extrapolated summary statistic, to populate the reference table;
    \item for each GP (one per summary), evaluated at $n_o$, we use the posterior mean and variance of the GP, as well as the empirical correlation between the summaries, to populate the reference table; these estimates are used to reconstruct the multivariate (here, bivariate) normal distribution of the summaries.
\end{enumerate}

In order to obtain the ABC posteriors, we need to compute some measures of similarity between the extrapolated and observed network statistics. For situation (i), we can use the same standardized Euclidean distance as for the least squares approach. We denote the associated results as GPc-ABC hereafter. For (ii), the reconstructed distributions at $n_o$ provide a natural way to do this by evaluating the bivariate normal density functions at the observed statistics.
Hence, we define the ABC posterior as the parameter values associated with generated networks providing the highest densities to the observed statistics. This strategy is denoted GPa-ABC hereafter. While this approach takes into consideration the extra information provided by the GP marginal distributions, one major problem can occur, which is the situation where all the extrapolated normal distributions give a zero density to the observed summaries. In theory this should not happen, however in practice this is possible due to numerical approximations, especially when the GP posterior variances are small. This is not an uncommon situation as the GP variances are usually lower a posteriori than a priori.
To prevent this issue we also display the GP results using (ii), but inflating the variance-covariance matrices by a factor $100$. This strategy is denoted GPb-ABC. 
An alternative strategy to benefit from the GP marginal distributions is to marginally evaluate each normal density at the observed summary statistic value and use, for example, the sum of these evaluated densities as a similarity measure between observed and extrapolated summary statistics. Because this methodology does not consider the correlation among summary statistics, we focused our investigations on GPa-ABC.

\section{Simulation Studies}
\label{simulation}

\begin{table}
\centerline{
\resizebox{0.9\columnwidth}{!}{%
\begin{tabular}{l|l|l|l}
Method      & Summary nature & Posteriors from & Other features\\
\hline 
S-ABC       & true                              & lowest distance  & \\\hline
LS-ABC      & ext. with least-squares           & lowest distance  & \\\hline
GPa-ABC     & ext. with GP means                & highest density & reconstructed normal \\\hline
GPb-ABC     & ext. with GP means                & highest density & reconstructed normal with\\
    &   &   & inflated covariance matrix\\\hline
GPc-ABC     & ext. with GP means                & lowest distance  & \\\hline
RE-ABC 500  & ext. with least-squares           & lowest distance  & sample triangle count tracked \\
  & & & up to $n_s=500$ nodes\\\hline
RE-ABC 1000  & ext. with least-squares          & lowest distance  & sample triangle count tracked \\
  & & & up to $n_s=1000$ nodes\\
\end{tabular}%
}
}
\caption{Summary of the different extrapolation method names and their features (``ext.'' stands for extrapolated).}
\label{tab:summary-method}
\end{table}

We conduct simulation studies to assess the performance of our extrapolation ABC procedure for the DMC model described above.
To facilitate the reading of the various tables and graphs present in the remainder of this manuscript, Table \ref{tab:summary-method} summarizes the different method abbreviations and their features.
The prior distributions of the DMC parameters are uniform over the rectangular parameter space with $q_{m}\in[0.15,0.35]$ and $q_{c}\in[0.1,0.9]$.
The reason for the tight restriction on the parameter space in terms of $q_{m}$ is that the functional form for the growth of average degree depends highly on this parameter in similar models \citep{ispolatov2005duplication}. For this proof of concept, we have thus restricted the parameter space to a subset of parameter values that result in polynomial growth. As in any Bayesian approach, the priors should be determined cautiously by practitioners.
Nonetheless, if one wants to increase the prior range, a naive strategy would be to choose an extrapolation method that is specific to a given part of the prior space, depending on the shape of the tracked summaries. This approach would however be tedious and instead we suggest employing a more flexible extrapolation method. Adapting Gaussian processes would involve either using a flexible family of kernels, for example the spectral mixture kernel \citep{lloyd:etal:2014} or using automatic combinations of different simple kernels \citep[see e.g.][]{wilson:adams:2013, sun:etal:2018}.

For each parameter sample drawn from the prior, a corresponding network realization is generated with average degree and number of triangles tracked every $5$ nodes from $35$ up to $n_s=500$ as the network grows.
In this way, we generate $B=4000$ realizations to populate our reference table.
As mentioned earlier, the growing process starts from a small seed network and its choice is not without influence on the final network.
This seed is often selected to reproduce certain characteristics of observed networks, and/or is motivated by specialist knowledge \citep[][]{hormozdiari:etal:2007, schweiger:etal:2011}. For example, when studying protein-protein-interaction (PPI) networks, \citet{hormozdiari:etal:2007} build a seed network made of $50$ nodes in order to reproduce cliques and normalized degree distribution of an observed yeast PPI network. 
In our simulation study, for simplicity, we use an Erdos–Rényi graph \citep[ER,][]{erdos:renyi:1960} of $30$ nodes with the probability parameter $p=0.2$; we also specify a random seed for the ER network generation function such that the resulting graph has one connected component only.
As a reminder, the ER model starts from a fixed number of unconnected nodes, and an edge between a node pair is turned on with probability $p$. We could also have started from a smaller graph, such as a triangle, however in this ABC framework a larger seed reduces the variability across simulated graphs given the same parameter values, and thus also reduces variability among the same summary statistic.

The two methods of extrapolation, as described above, are based on the tracked quantities, average degree and number of triangles, for various values of $n_o=1000$, $2000$, $5000$, $10000$. For each $n_o$, we grow $200$ networks up to $n_o$ nodes for each pair of parameter values $(q_{m},q_{c})\in\{(0.2, 0.3)$, $(0.2, 0.7)$, $(0.25, 0.5)$, $(0.3, 0.2)$, $(0.3, 0.7)\}$ in the interior of the parameter space. Each set of $200$ replicate simulations is used as test (observed) data to assess the ability of the methods to retrieve the corresponding true pair of parameter values. The resulting ABC posterior for each method of extrapolation is compared against the classical ABC posterior obtained by generating networks fully up to $n_o$ nodes and the posterior mean is averaged across $200$ ABC posteriors. We similarly estimate the standard deviation (SD) and root mean squared error (RMSE) of the posterior means. For all methods relying on a standardized Euclidean distance, to form the ABC posterior samples, we retain the simulated parameters achieving the $50$ smallest distances. For the Gaussian process extrapolation using the reconstruction of the bivariate normal distributions, in order to keep the results comparable, the ABC posterior is defined as the parameters providing the $50$ highest density values. Analysis based on non-extrapolated summaries is termed here \textit{standard} ABC (S-ABC), as it provides a more conventional contrast over extrapolated summaries.

When appropriate, the standardized Euclidean distance is based on the standard deviation of each summary. We approximate these values at $n_o$, using a set of $1000$ simulated networks built up to $n_o$.
Even though this requires additional computation, it is well known that this standardization is crucial for most ABC methods \citep[see e.g.][]{prangle:2017}.
One way to circumvent this is to instead use the extrapolated summary values. Both strategies lead to comparable results as long as the extrapolation methods lead to similar standardization values. Nonetheless, by using non-extrapolated summaries we can interpret large differences between ABC posteriors as solely due to the extrapolation quality, without dependence to the standardization used.

\begin{figure}
\centerline{
\includegraphics[width=1.05\linewidth, trim={3.7cm 2.78cm 3.5cm 5cm},clip]{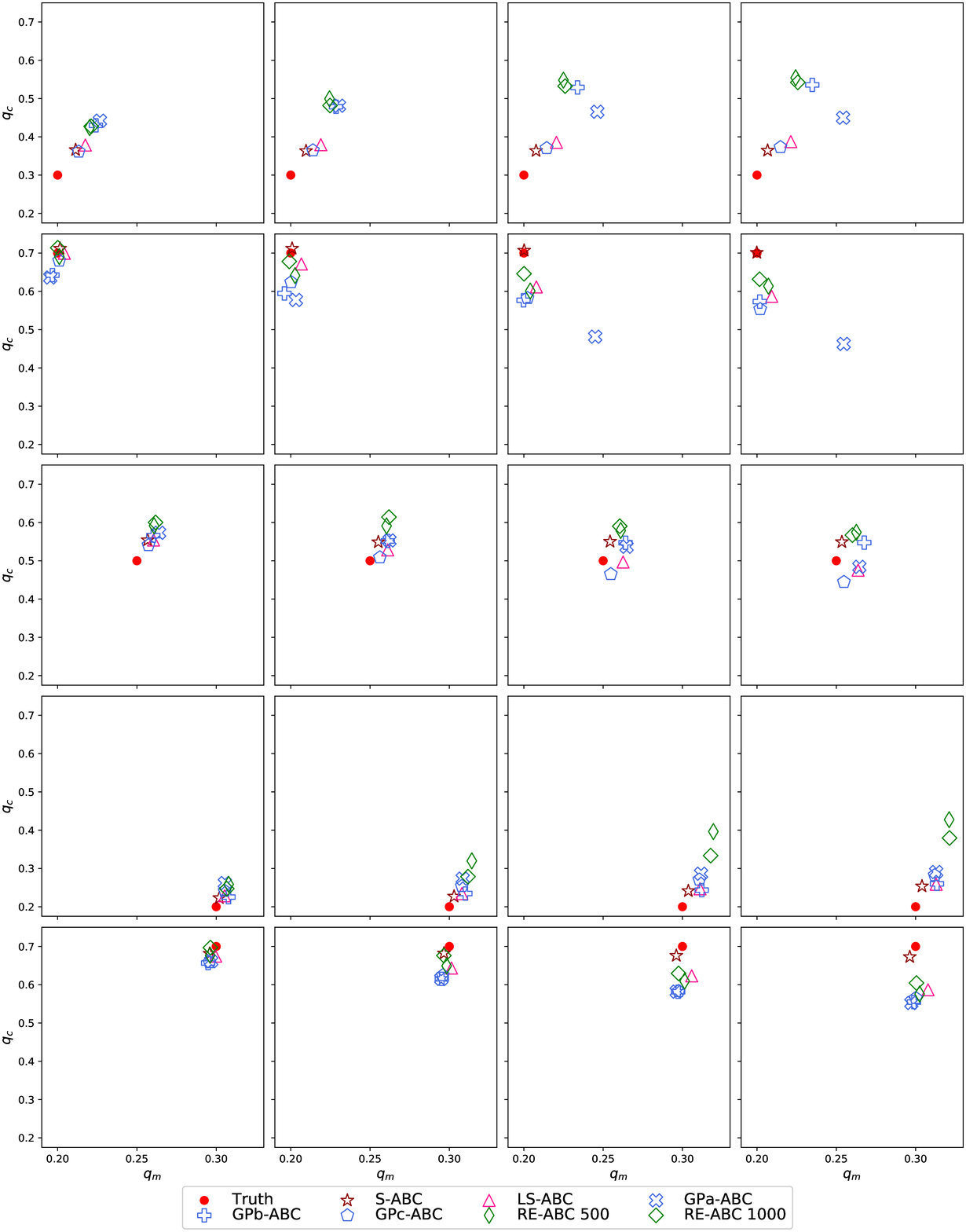}
}
\caption{Average posterior means provided by the various methods, with the red dot denoting the true parameter value. Each row corresponds to different values of the true parameter $(q_{m},q_{c})\in\{(0.2, 0.3)$, $(0.2, 0.7)$, $(0.25, 0.5)$, $(0.3, 0.2)$, $(0.3, 0.7)\}$, each column corresponds (from left to right) to values of $n_o=$ $1000$, $2000$, $5000$, $10000$.}
\label{fig:DMC:point_estimate_loc}
\end{figure}

Figure \ref{fig:DMC:point_estimate_loc} summarizes the posterior means for each combination of true parameter value and $n_o$. In general, across all combinations, there seems to be more significant bias in the estimate of $q_{c}$ regardless of the method used to obtain the ABC posterior.
We observe that the average S-ABC posterior means are relatively unchanged with the number of nodes, while posteriors based on extrapolated summaries are further from the truth the larger $n_o$ is, but are still very similar to the standard ABC for $1000$ and $2000$ nodes.
The LS-ABC and GPc-ABC posteriors are often the closest to the standard ABC across all true parameter pairs. 
The similarity between these two extrapolation methods is not surprising as the least squares function is provided as GP prior mean function.
For GPa-ABC, on the two first parameters, we observe a characteristic bias toward the prior mean $(0.5, 0.25)$ when $n_o$ increases. This is explained by the numerical problem mentioned earlier, namely that less than $50$ reconstructed normal distributions are able to give non-zero density to the observed summaries, meaning that simulated parameters are instead randomly selected over the prior space. This bias increases with $n_o$ for two reasons: first, the normal distributions might be centered in regions of the summary space too far from the observations, this is accentuated by extrapolation errors that can occur for a larger $n_o$; second, the variance-covariance matrices induce too peaked normal distributions. It is to prevent this second case that GPb-ABC artificially inflates these matrices by a factor of $100$. Note that a higher factor did not change results. 
Another way to reduce the impact of the numerical problem of having zero densities would have been to increase the total number of simulated data points, which would have improved the performance of every method.
As expected, the GPb-ABC strategy manages to reduce the bias for the problematic parameters, however results remain less satisfactory than when using the GP distance-based strategy (GPc-ABC).
Finally, for the third pair of parameters, $(0.25, 0.5)$, all average posterior means are close to the truth, which, for this specific case, coincides with the prior mean. Because of this coincidence, it is difficult to determine whether or not the performance of GPa-ABC is due to its bias toward the prior mean.

\begin{figure}
\centering
\includegraphics[width=\linewidth, trim={3.25cm 3cm 3.5cm 5cm},clip]{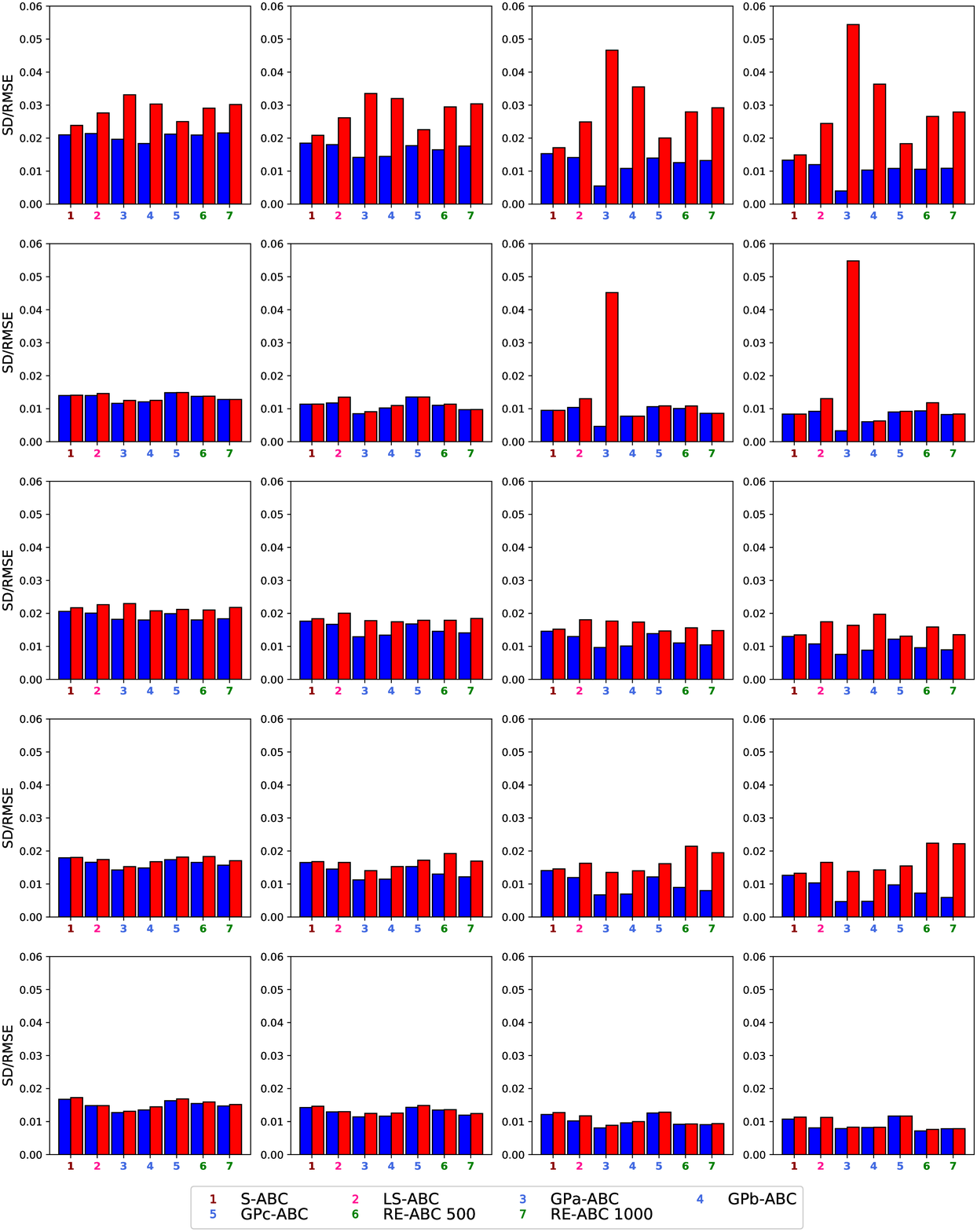}
\caption{Standard deviation (SD, left bar of each pair) and root mean square error (RMSE, right bar of each pair) of the different estimators for $q_{m}$.
Each row corresponds to different values of the true parameter $(q_{m},q_{c})\in\{(0.2, 0.3)$, $(0.2, 0.7)$, $(0.25, 0.5)$, $(0.3, 0.2)$, $(0.3, 0.7)\}$, each column corresponds (from left to right)  to values of $n_o=$ $1000$, $2000$, $5000$, $10000$.}
\label{fig:DMC:sd_rmse_qm}
\end{figure}

\begin{figure}
\centering
\includegraphics[width=\linewidth, trim={3.25cm 3cm 3.5cm 5cm},clip]{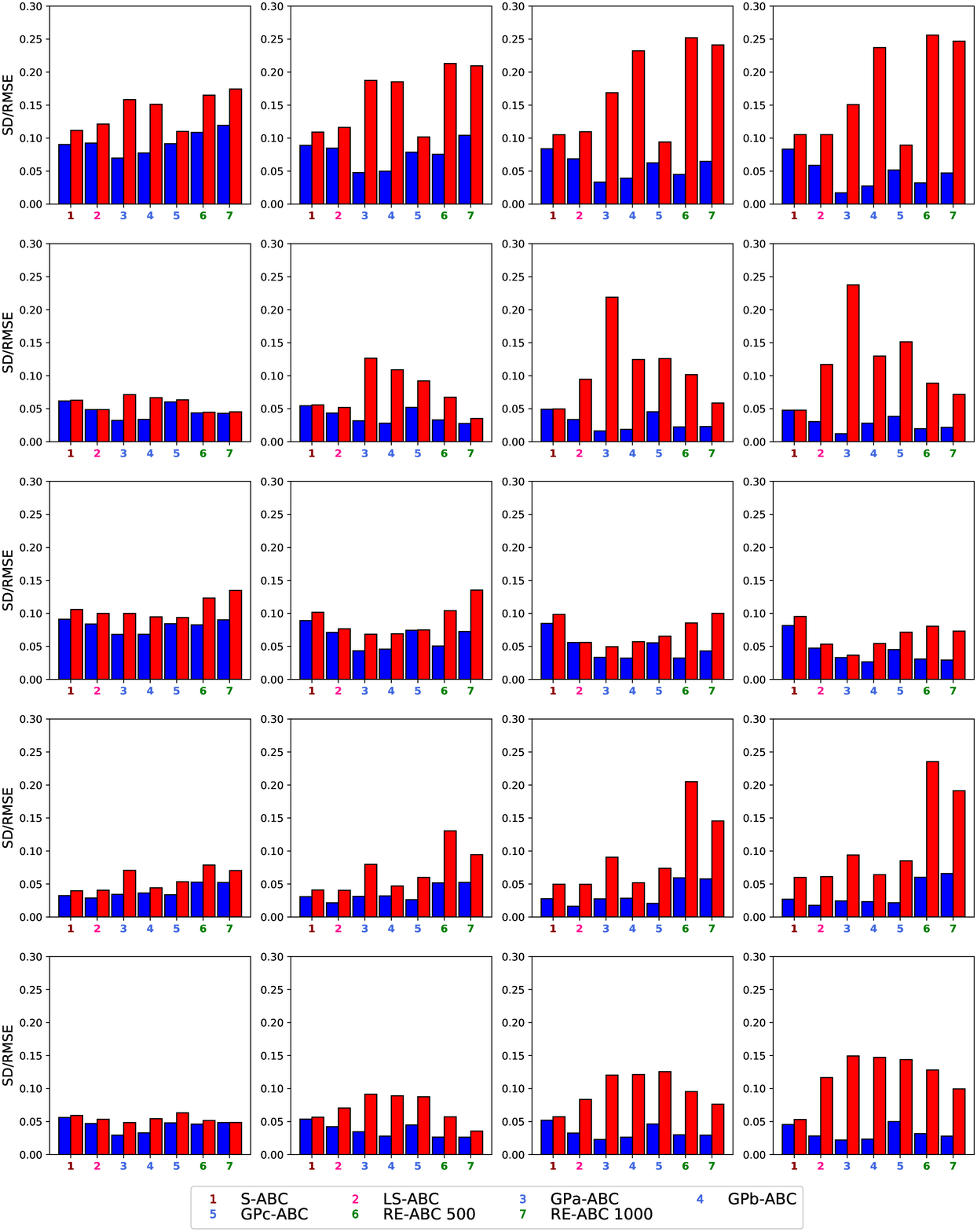}
\caption{Standard deviation (SD, left bar of each pair) and root mean square error (RMSE, right bar of each pair) of the different estimators for $q_{c}$.
Each row corresponds to different values of the true parameter $(q_{m},q_{c})\in\{(0.2, 0.3)$, $(0.2, 0.7)$, $(0.25, 0.5)$, $(0.3, 0.2)$, $(0.3, 0.7)\}$, each column corresponds (from left to right) to values of $n_o=$ $1000$, $2000$, $5000$, $10000$.}
\label{fig:DMC:sd_rmse_qc}
\end{figure}

Figures \ref{fig:DMC:sd_rmse_qm} and \ref{fig:DMC:sd_rmse_qc} summarize for each true parameter, the standard deviation (SD) (left bar of each pair) of the $200$ posterior means, and root mean square error (RMSE) (right bar of each pair) between the posterior means and the true parameter value, for each combination. Most of the comments emerging from Figure \ref{fig:DMC:point_estimate_loc} are still visible here. Regarding the parameter $q_m$, LS-ABC and GPc-ABC provide the most similar SD and RMSE compared to the standard ABC. Both methods are also showing high robustness to the growth of $n_o$. For $q_c$ however, the degradation with this growth can be observed. Concerning the GP density-based method, GPa-ABC, the SD is characteristically very low, because most of the posteriors are centered at the prior mean.

Given the above results, there seems to be potential in the extrapolation methods, however as one might expect the extrapolation quality is pivotal here. Moreover, this quality is also parameter dependent. When a moderate number of nodes is observed, the similarities in performance of LS-ABC and GPc-ABC compared to S-ABC suggest that, at the bare minimum, one should use one of these two first as their bias from the standard ABC is low, while the gains in computation time are substantial.
Contrary to the least squares, the GP still has room for improvement: it has the potential to be refined in problem-specific ways, but also more general/automatic kernels could be used to improve extrapolation quality across the different regions of the parameter space, see for example \citet{wilson:adams:2013, lloyd:etal:2014, sun:etal:2018}.

\section{Subsampling Computationally Intensive Summaries}
\label{subsampling}

In addition to forward simulation from models, ABC requires evaluation of summary statistics. Because generating some network summaries is computationally expensive, another opportunity to save computational resources is to base these summaries on a subsample of the network rather than the whole network. While the computation of average degree scales well with network size $n$, a more complex summary such as triangle enumeration is trivially solvable in $O(n^3)$ and the best-known algorithm takes time $O(n^{2.373})$ on sparse power-law graphs using fast matrix multiplication \citep{latapy2008main}.
As a potential alternative for better scaling, as the network grows, we propose to count the number of triangles in a subgraph induced by a set of randomly sampled nodes without replacement. We call this the \emph{sample triangle count} and contrast it with the \emph{population triangle count} computed over the whole population of nodes (the whole network).

\begin{figure}
\centering
\includegraphics[width=0.9\linewidth, trim={1.25cm 0.25cm 1.25cm 1.25cm},clip]{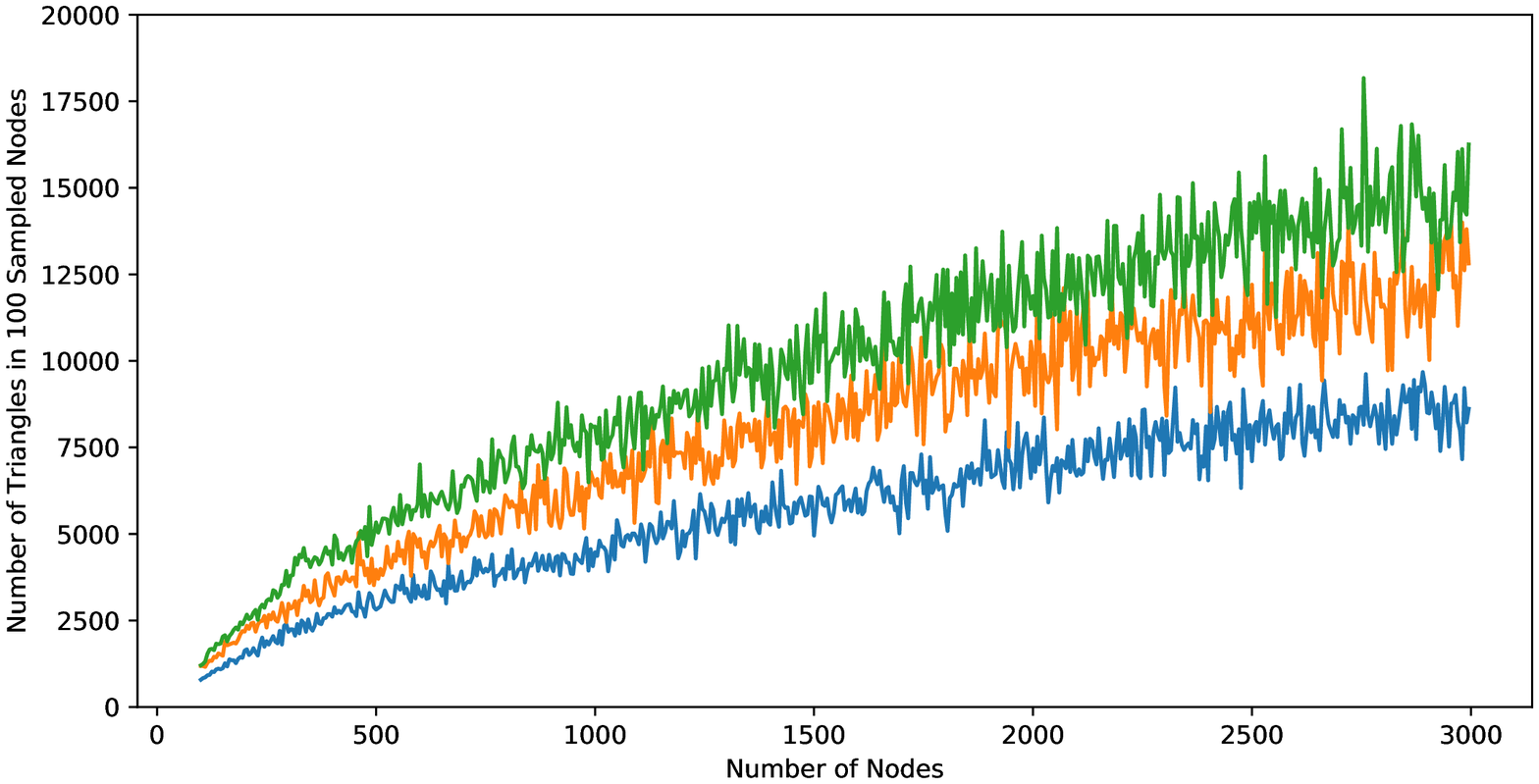}
\caption{Growth of the number of triangles in a random sample of 100 nodes from network realizations generated from the DMC model.}
\label{fig:DMC:subsampled_triangles}
\end{figure}

Similarly to Figure \ref{fig:DMC:summary_sim}, Figure \ref{fig:DMC:subsampled_triangles} shows the number of triangles in 100 randomly sampled nodes for three network realizations generated from a DMC model.
While this summary shows much greater variability due to the sampling, the growth pattern is approximately polynomial. We therefore fit the sample triangle count to the function $\tilde{s}_b(n) = a_b  n^{c_b} + d_b$. The number of triangles in the observed network generated from the true parameter values will also be based on $100$ subsampled nodes, meaning that we use the same summary statistic for the observed network.
Outside of the node samplings and the use of this different polynomial function, we generated $1000$ networks up to $n_o$ nodes, with parameters sampled from the prior, to determine the empirical standard deviation for the sample triangle obtained on them. This empirical standard deviation is needed to compute the Euclidean distances. Everything else is the same as in least squares ABC, including the extrapolation. We will refer to this approach as RE-ABC.

\begin{figure}
\centering
\includegraphics[width=\linewidth, trim={3cm 0.5cm 3.5cm 1cm},clip]{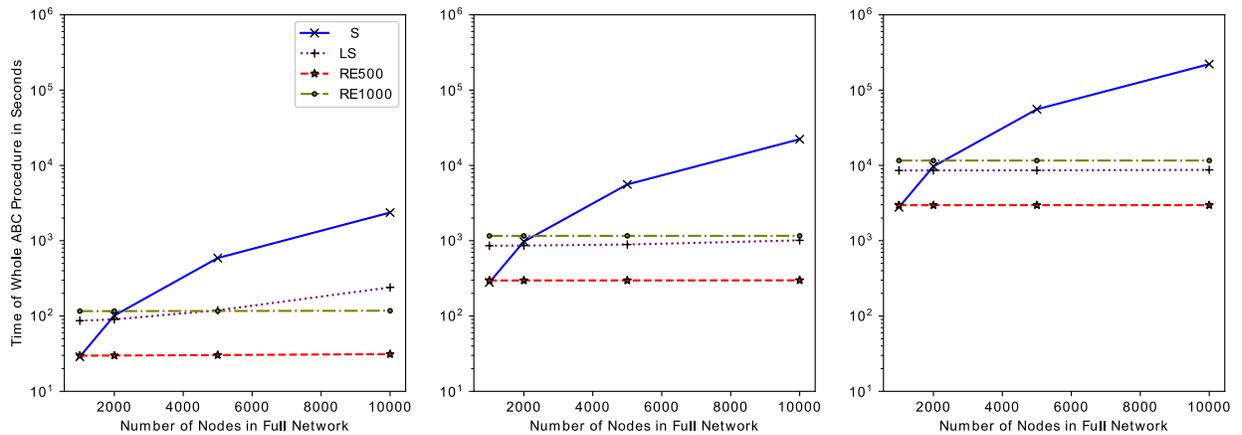}
\caption{The time required for the entire ABC procedure for various sizes of the full network for S-ABC (blue), LS-ABC (purple), RE-ABC to $n_s=500$ (red), RE-ABC to $n_s=1000$ (olive)
for a reference table of size $10$, $100$, $1000$ (from left to right).}
\label{fig:DMC:total_times}
\end{figure}

Before assessing performance, we assess the time required for each method, both for constructing one entry of the reference table, which entails generating the network and tracking the statistics when applicable, and time needed to compute the statistic for the observed network, which is required only once per procedure. For RE-ABC, we consider tracking the statistics up to both $n_s=500$ and $1000$ but randomly sample $100$ nodes in both cases for the triangle count, while for LS-ABC, we only tracked up to $n_s=500$. 
We include this greater value for $n_s$ because it is less expensive computationally as it based on a sample triangle count instead of the population triangle count.
The results averaged over $100$ replications are reported in Tables \ref{tab:DMC:single_entry_times} and \ref{tab:DMC:full_subsampled_times}.
Note that the time to construct a single entry in the reference table is independent of $n_o$ for LS-ABC and RE-ABC since
the tracking is only for $n_1,\ldots, n_s$ regardless of $n_o$.
While S-ABC is the fastest for the smallest value of $n_o$, the other methods quickly overtake S-ABC as $n_o$ increases.
As expected, it is consistently faster to compute the network statistics for the subsample than for the full observed network, the speed-up being greater for large $n_o$. As a more holistic measure of time, we compare the time needed to build a reference table of various fixed sizes for each method as well as the time required to compute the statistics for the observed network once. The results are summarized in Figure \ref{fig:DMC:total_times}. Using RE-ABC tracked up to $n_s=500$ is consistently the fastest, except for $n_o=1000$,
while S-ABC is consistently the slowest for $n_o=5000$ and $10000$. For $n_o=2000$, S-ABC, LS-ABC, and RE-ABC tracked up to $n_s=1000$ all have similar computation time. Between LS-ABC and RE-ABC tracked up to $n_s=1000$,
the ordering of the time required can depend on the size of the reference table. Regardless, the proposed methods all offer significant speed-ups compared to S-ABC except for small values of $n_o$. The time difference between LS-ABC and RE-ABC depends on the size of the reference table, and
the value of $n_s$ and $n_o$. While the numbers here may not always be realistic in practice, this example illustrates the improvements in computation time offered by the proposed methods. In the next section, we assess the time required for a large empirical network.

\begin{table}
\centerline{
\begin{tabular}{c|cccc}
Method \textbackslash \, $n_o$ & 1000 & 2000 & 5000 & 10000\\
\hline 
S-ABC & 2.781 &  9.711 & 55.609 & 221.240\\
LS-ABC & 8.572 & 8.572 & 8.572 & 8.572\\
RE-ABC 500 & 2.966 & 2.966 & 2.966 & 2.966\\
RE-ABC 1000 & 11.599 & 11.599 & 11.599 & 11.599\\
\end{tabular}}
\caption{Average time (in seconds) to construct a single entry in the ABC reference table for the various methods.}
\label{tab:DMC:single_entry_times}
\end{table}

\begin{table}
\centerline{
\begin{tabular}{c|cccc}
Summary type \textbackslash \, $n_o$ & 1000 & 2000 & 5000 & 10000\\
\hline 
Full & 1.003 &  4.488 &  33.387 & 154.484\\
Subsampled & 0.101 & 0.226 & 0.666 & 1.567\\
\end{tabular}}
\caption{Average time (in seconds) to compute statistic for full network and subsampled nodes.}
\label{tab:DMC:full_subsampled_times}
\end{table}

Figures \ref{fig:DMC:point_estimate_loc}, \ref{fig:DMC:sd_rmse_qm} and \ref{fig:DMC:sd_rmse_qc} show the bias as well as SD and RMSE of RE-ABC, tracked up to $n_s=500$ and $1000$, in comparison with the other methods.
RE-ABC exhibits the same behavior as all other extrapolation methods, namely that the ABC posterior accuracy diminishes when the range of extrapolation increases. Tracking sample-based summaries up to $n_s=1000$ leads to better results than up to $500$, which suggests a quicker deterioration in the quality of the extrapolated quantities when using smaller $n_s$ values, as expected.
Interestingly, when $n_s = n_o = 1000$,  and in general when the number of tracked nodes coincides with the size of the observed network, for the second and fifth parameter pairs (i.e., for some but not all parameter combinations), RE-ABC can lead to even better results than S-ABC, suggesting that the sample triangle count can be as informative as the population triangle count. 
It is important to note that at $n_o = 1000$ we still use the fitted polynomial functions to determine the summary statistic values for this number of nodes rather than using the raw sample triangle count for $n_s=1000$, which is much more noisy.

This example shows the potential decrease in computation time for RE-ABC, but also the potential deterioration in performance that depends on the value of $n_s$ as well as the extrapolation range, i.e. $n_o$. Nonetheless, posterior means are comparable to results obtained thanks to population summaries. As a final remark, as shown in Figure \ref{fig:DMC:subsampled_triangles}, this type of summary presents large subsampling variance across nodes, and to reduce this variability, an intuitive idea is to draw the sample triangle count multiple times, and use the average of these quantities as our summary.
As an example, given a randomly generated DMC network with 1000 nodes, the empirical variance of the sample triangle count is equal to 174,099. It drops to 15,232 and 8,308 when computing the average sample triangle count over 10 or 20 replicates of the subsampling procedure, respectively.
We performed extrapolations based on such a summary, and while the variability is highly reduced, for this example the least squares fitted functions were barely unchanged compared to the non-averaged version, leading to almost identical ABC posterior means (Appendix \ref{appendix:subsec:A.2}). Nonetheless, this consideration is further investigated in Section \ref{empirical}, where a single observed network is studied.

\section{Application to Scientific Citation Networks}
\label{empirical}

We apply the proposed procedure to an empirical citation network from the American Physical Society \citep{APSdata}, containing $n_o=$ 597,819 articles starting $1893$ with over $7$ million citations (corresponding to directed edges). 
Even though more complex mechanistic models could be employed, we make use of the Price model \citep{price1965networks}, where each newly added node attaches itself, on average, to $m$ existing nodes such that the probability to attach to a given node is proportional to $k_0+k$, where $k_0$ is a constant and $k$ is the in-degree of the node. The number of nodes a new node attaches to is generated from a binomial distribution $B(610,p)$, where the upper bound of $610$ is motivated by the maximum out-degree in the empirical network. The prior distribution of $(k_0,p)$ is uniform over the rectangle $[0.9,1.1]\times[0.019,0.021]$. The center of the prior for $k_0$ is based on Price's original proposal, while that for $p$ is obtained by matching the first moment of the empirical network's out-degree distribution.

We drew $400$ parameter samples from the prior and generated network realizations up to $n_o=$ 597,819 nodes from a seed network of 28,645 nodes consisting of papers published before $1960$ inclusively.
The LS-ABC (only extrapolation) and RE-ABC (subsampling and extrapolation) results are based on the mean and variance of the in-degree distribution and the number of triangles tracked for 50,000 nodes, i.e., $n_s=$ 28,645 $+$ 50,000, at every 50 nodes. Of the three tracked statistics, the number of triangles is the most computationally intensive and is based on a subsample of 25,000 nodes for RE-ABC. We also investigated the influence of reducing the sampling variability by considering the averaged sample triangle count obtained over 10 or 20 replicates (denoted RE-ABC 10 avg and RE-ABC 20 avg). The S-ABC results are based on the three statistics computed at the full network size of $n_o$ nodes. The ABC posterior distributions are deduced in the same way as in the previous example, and for the Euclidean distance computation, we used $100$ simulated networks to determine the required empirical standard deviation of the summaries at $n_o$.

The mean of the in-degree distribution can be seen as a weighted average between the mean of the in-degree distribution of the seed network and contributions from
newly added nodes which give a mean of $m$ new in-edges. This motivates a functional form of $\tilde{s}_{b}(n)=a_b/n+c_b$ for the mean of the in-degree distribution for LS-ABC. We also know that the in-degree distribution is approximately of power-law form $\P(k_i=k)\sim k^{-\alpha}$.
Thus, the second moment can be seen as a generalized harmonic number, $\E(k^2)\sim \sum_{k=1}^{k_{\text{max}}}k^{2-\alpha}$.
Note that this summation is not infinite, since we are interested in the variance of the in-degree distribution of networks of a given size, and they must have an upper bound on the in-degree. The harmonic number $H_{k_{\text{max}}}=\sum_{k=1}^{k_{\text{max}}}\frac{1}{k}$ has analytical
form $\gamma+\psi_0(k_{\text{max}}+1)$, where $\gamma$ is the Euler-Mascheroni constant and $\psi_0$ is the digamma function. This motivates the functional form we use to extrapolate the variance of the in-degree distribution as
$\tilde{s}_{b}(n)=(\gamma+\psi_0(a_b n + 1))^{c_b}+d_b$. Finally, for triangles, we employ a polynomial functional form, $\tilde{s}_{b}(n)= a_b n^{c_b}+d_b$, for both LS-ABC and RE-ABC.

\begin{figure}
\centering
\includegraphics[width=\linewidth, trim={1.75cm 2.5cm 3cm 3cm},clip]{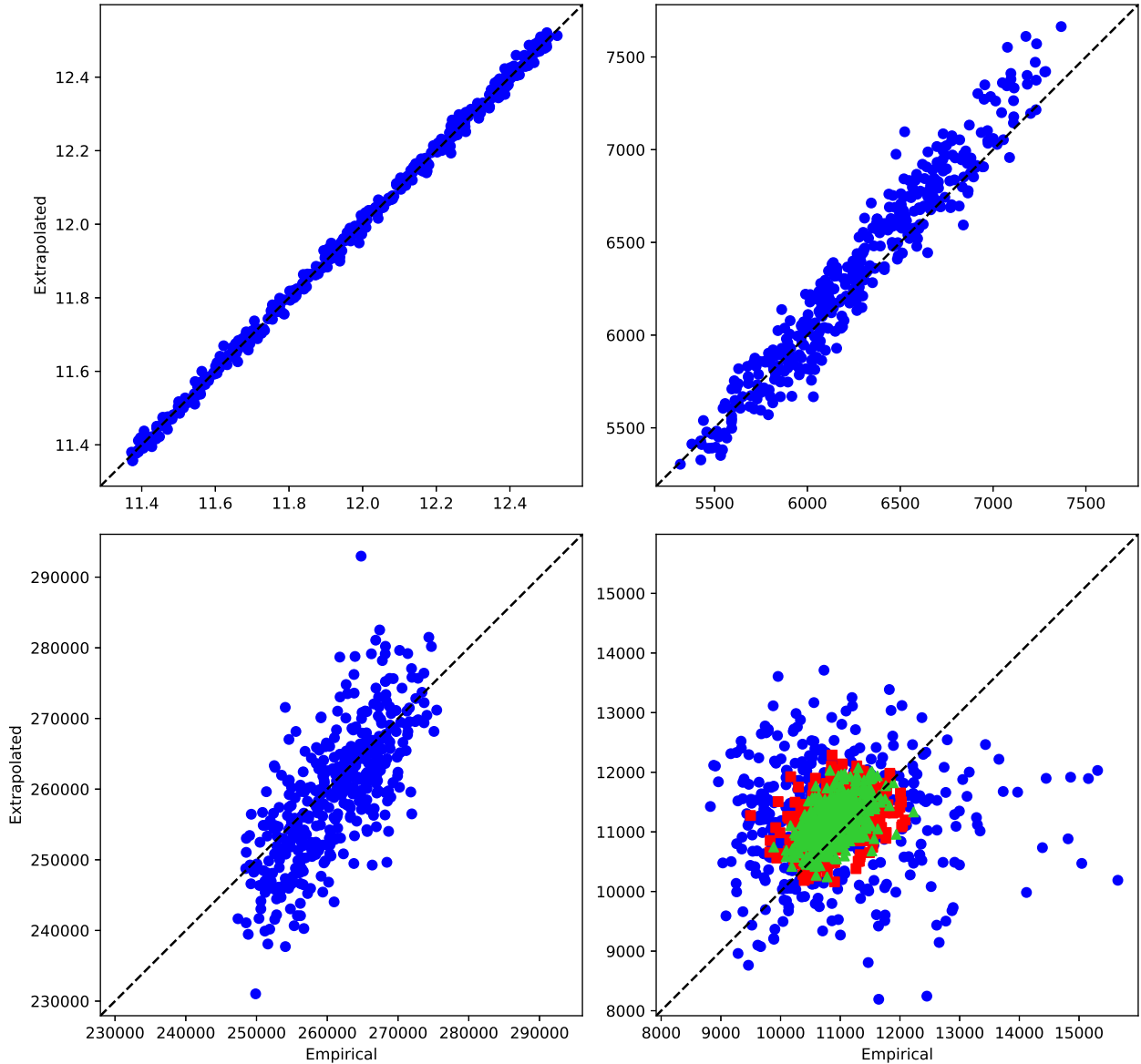}
\caption{The empirical vs.~the extrapolated value of the mean (top left) and variance (top right) of the in-degree distribution, number of triangles without sampling (bottom left), and number of triangles with sampling (bottom right). On the  bottom right graph, blue circles represent $1$ sample triangle count, while red squares and green triangles correspond to the averaged sample triangle count over $10$ and $20$ independent replicates of 25,000 nodes respectively.}
\label{fig:citation:extrapolation_quality}
\end{figure}

The empirical value ($x$-axis) measured at $n_o$ versus the extrapolated value ($y$-axis) of the summary statistics for each generated network realization is shown in Figure \ref{fig:citation:extrapolation_quality}.
The mean and variance of the in-degree distribution are well extrapolated. While the extrapolated population triangle count does not perform quite as well compared to the mean and variance, it is still close to the true count, despite the slightly larger spread around the identity line. For the sample triangle count used by RE-ABC, when using a single subsampling procedure, the points are still positioned around the identity line, i.e., no bias is induced. This is also the case when using an average over 10 or 20 replicates of the subsampling procedure.
The non-averaged version exhibits the highest spread, while the averaged summaries yield more accurate extrapolated values when compared with the empirical values. By reducing subsampling variability, we generate more accurate extrapolated summaries, and we therefore expect more accurate posterior distributions when using the averaged summaries. The cost of greater precision here is the added computational load needed for the replications.
We also notice a slight improvement when using 20 replicates instead of 10.

\begin{figure}
\centering
\includegraphics[width=0.95\linewidth, trim={3.25cm 0.25cm 3.5cm 1.5cm},clip]{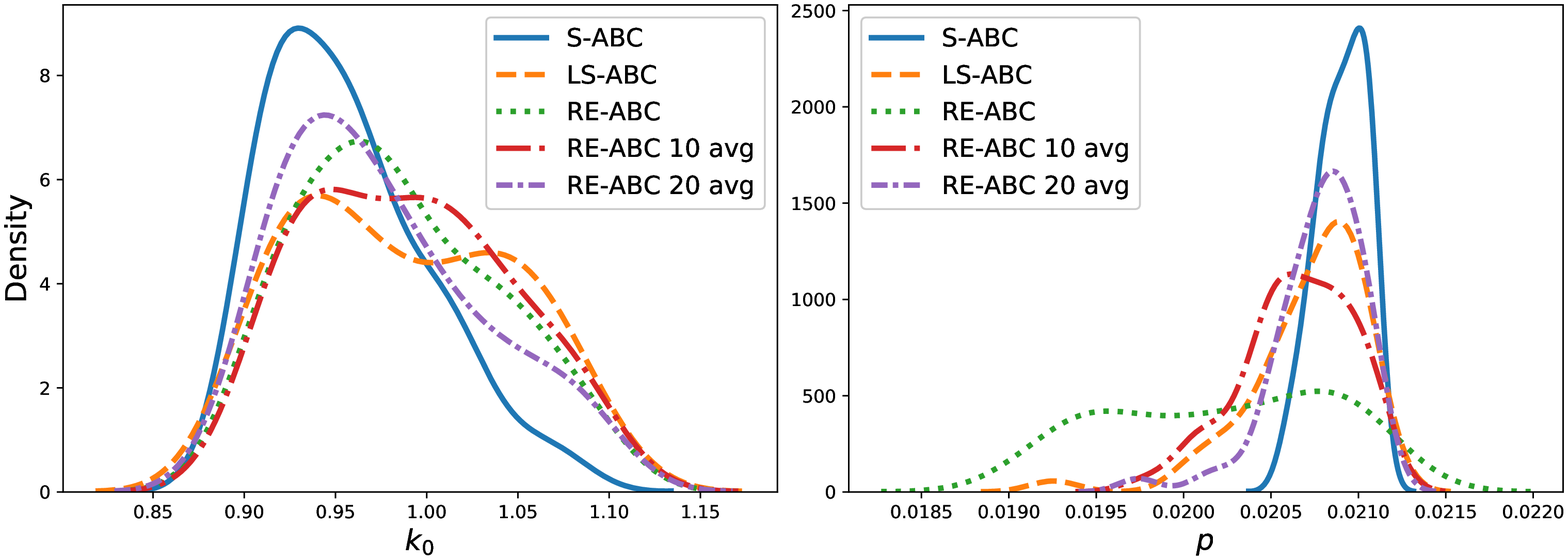}
\includegraphics[width=0.95\linewidth, trim={3.25cm 0.25cm 3.5cm 1.5cm},clip]{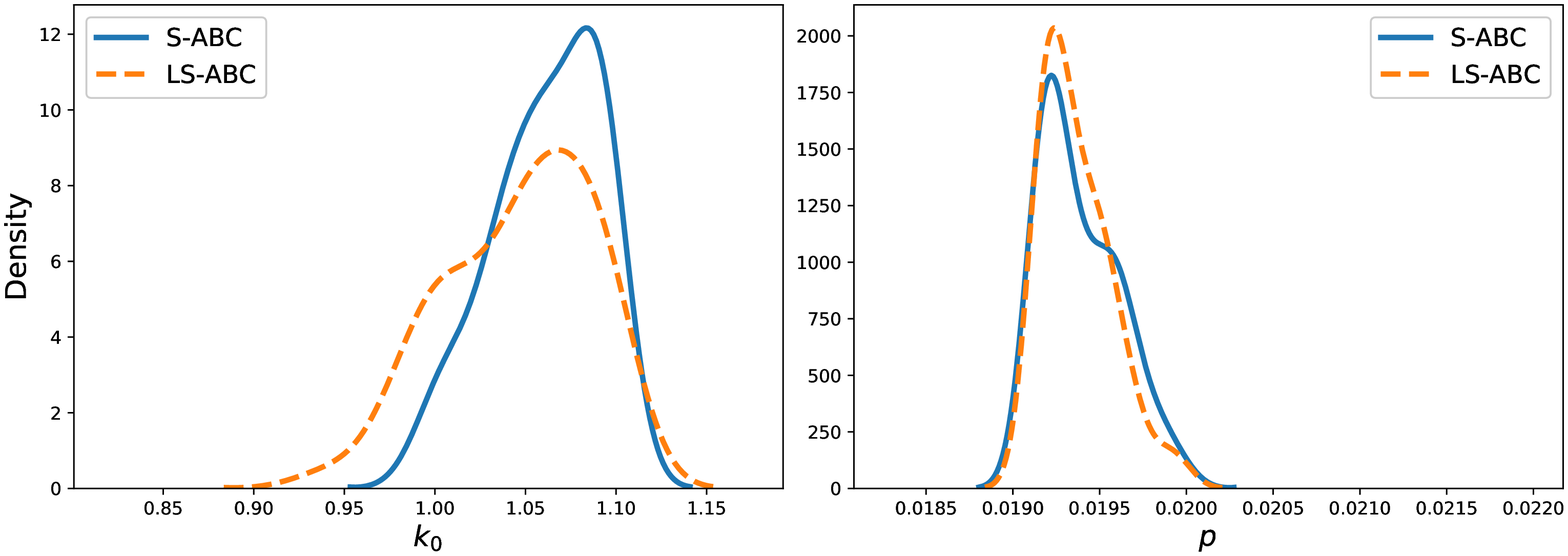}
\caption{Posterior distributions for the different methods. The first row includes the triangle count as summary, while the second row excludes it.}
\label{fig:citation:posteriors_with_and_without_triangles}
\end{figure}

Figure \ref{fig:citation:posteriors_with_and_without_triangles} displays the estimated posterior distributions for the different methods, when including or not the triangle count in the summary statistics set, population or sample based when appropriate. Including or excluding this summary gives some insights on its influence on the final posteriors, especially since it has the worst extrapolation quality.
Table \ref{tab:citation:posterior_quantities_with_without_triangle} summarizes posterior quantities of interest for the different approaches.

When making use of the triangles, for the parameter $k_o$, we observe that all extrapolated posteriors are relatively close to the S-ABC. Using sample triangle count provides even better posteriors compared to LS-ABC, but no great improvement can be observed when replicating the sampling procedure 10 or 20 times. However, this is not true for the parameter $p$. Indeed, without using replicates, the RE-ABC posterior spreads over the prior range, while the posterior gets closer to S-ABC with a higher number of replicates. When examining Table \ref{tab:citation:posterior_quantities_with_without_triangle}, we notice that the posterior variances are always overestimated compared to S-ABC, and in general RE-ABC (20 avg) provides the most similar results.

If triangles are not used as a part of the ABC procedure, the S-ABC and LS-ABC posteriors match even more closely. This is especially encouraging for the second parameter.
In this example, our proposed ABC procedure works well, especially when the summary statistics are relevant and well extrapolated. While selection/relevance of summary statistics is important for ABC methods based on distance, a more detailed discussion of this question is not in the scope of our paper.

There is also significant difference between the time required for each procedure.
To compute each element of the ABC reference table on average, S-ABC takes 221,290 seconds, LS-ABC takes 13,047 seconds, and RE-ABC using 1, 10 and 20 sample triangle counts takes respectively 8,724, 38,835 and 73,997 seconds.
Computing the summary statistics on the full observed network (for S-ABC and LS-ABC) takes 236 seconds, while RE-ABC with 1 subsampling only needs 11 seconds. However, by using averaged sample-based summaries we broadly multiply this time by the number of used replicates, reaching 117 seconds when using 10, and 209 seconds for 20.
While LS-ABC reduces computation time, RE-ABC shows an even larger reduction when using a single sampling replicate, although using too many replicates will increase computation burden. The more time intensive summary statistic, triangle count, did not take too long to compute even on the full network, which here made our comparisons possible. However, it would not be feasible to compute summaries with similar or worse time complexity on very large networks for all data. The relative time reduction due to RE-ABC in computing both the elements of the reference table as well as the summary statistics of the full empirical network is significant. Even though using averaged sample-based summaries involves an increase in computation, the results might be close to the classic ABC posterior, as observed in this example, making it worth consideration.

Given that the data cover a fairly long time period, it is possible that the mechanisms responsible for the growth and evolution of the citation network change over time. A comprehensive way to address this question would be to consider a model selection problem for different time intervals, where different network models incorporate different growth mechanisms. We have considered the model selection problem for mechanistic network models elsewhere \citep[see e.g.][]{chen2018flexible}, and because this paper focuses on parameter inference, we follow a somewhat different approach. Here we fit the same canonical model of network growth to two overlapping time periods: the shorter period from 1950 up to 2017 and the longer period from 1960 up to 2017. The posterior means ($95\%$ credible intervals) for $k_0$ and $p$ for the shorter period are $0.94912$ $(0.90576,1.01612)$ and $0.02086$ $(0.02056,0.02108)$, respectively, whereas the corresponding estimates for the longer period are $0.95707$ $(0.90657,1.09731)$ and $0.02091$ $(0.02064,0.02108)$. These results were obtained using triangle count as a summary, and this similitude did not change appreciably when that summary was omitted (not shown). The similarity of the posterior point estimates and credible intervals for the two intervals suggests that the mechanisms responsible for the evolution of the citation network are stable or ``stationary'' over the 10-year period, or at the very least that a possible change in network mechanisms has no influence on the resulting ABC posterior distributions for the given model.

\begin{table}
\centerline{
\setlength{\tabcolsep}{5pt}
\renewcommand{\arraystretch}{1.1}
\resizebox{0.95\textwidth}{!}{%
\begin{tabular}{l|ccccc|cc}
                        & \multicolumn{5}{c|}{With triangles} & \multicolumn{2}{c}{Without triangles} \\ \cline{2-8}
                        & S-ABC & LS-ABC & RE-ABC & \begin{tabular}[c]{@{}c@{}}RE-ABC\\ 10 avg\end{tabular} & \begin{tabular}[c]{@{}c@{}}RE-ABC\\ 20 avg\end{tabular} & S-ABC & LS-ABC\\ \hline
$\hat\E(k_0\mid \textbf{S})$                     & $0.95707$ & $0.98684$ & $0.98414$ & $0.98918$ & $0.97615$ & $1.06176$ & $1.04600$\\
$\hat\E(p\mid \textbf{S})$                       & $0.02091$ & $0.02071$ & $0.02022$ & $0.02065$ & $0.02077$ & $0.01939$ & $0.01937$\\
$\hat\Var(k_0\mid \textbf{S})$                   & $0.00182$ & $0.00328$ & $0.00270$ & $0.00282$ & $0.00278$ & $0.00083$ & $0.00154$\\
$\hat\Var(p\mid \textbf{S})\,(\times 10^{-7})$   & $0.18944$ & $1.16371$ & $3.83804$ & $0.98307$ & $0.65569$ & $0.48474$ & $0.38601$\\
$\hat\Q_{2.5\%}(k_0\mid \textbf{S})$             & $0.90656$ & $0.90850$ & $0.91155$ & $0.91289$ & $0.90909$ & $1.00105$ & $0.97723$\\
$\hat\Q_{2.5\%}(p\mid \textbf{S})$               & $0.02064$ & $0.02003$ & $0.01920$ & $0.02003$ & $0.02017$ & $0.01913$ & $0.01913$\\
$\hat\Q_{97.5\%}(k_0\mid \textbf{S})$            & $1.05689$ & $1.08409$ & $1.08466$ & $1.08444$ & $1.08430$ & $1.09731$ & $1.09731$\\
$\hat\Q_{97.5\%}(p\mid \textbf{S})$              & $0.02108$ & $0.02108$ & $0.02105$ & $0.02108$ & $0.02108$ & $0.01986$ & $0.01984$
\end{tabular}}}
\caption{Approximated posterior mean, variance, $2.5\%$ and $97.5\%$ order quantiles for the two parameters, obtained for the different approaches when including or not the summaries based on triangles.}
\label{tab:citation:posterior_quantities_with_without_triangle}
\end{table}

\section{Discussion}
\label{end}
Application of ABC to inference and model selection problems for mechanistic network models requires (i) forward simulation of network realizations from the given model and given parameter value, and (ii) computation of informative summary statistics that characterize the simulated network. Either or both of these steps may be computationally expensive. We proposed two methodological developments to make ABC feasible for modeling large networks using mechanistic models. First, because most mechanistic models grow the network starting from a small seed network, we proposed growing the network to a size smaller than the actual size of the observed network $n_o$ and then using extrapolation to estimate the values of the summary statistics at $n_o$. Second, we proposed the use of sample-based summary statistics (possibly averaged over multiple replicates), so that rather than computing the summary for the whole nodes in the simulated network, we only compute it on the graph formed by a subset of them. This approach works well for relatively local network summaries, such as degree or triangle count, but requires care when applied to more global network metrics since these metrics may exhibit high sampling variability. Although we used mechanistic network models to demonstrate this approach, both extrapolated summaries and sampled summaries are expected to be relevant in other ABC settings where the data are generated incrementally.

One important aspect of the results is that the target of our extrapolation-based ABC procedure is always the standard ABC (S-ABC) rather than the true parameter value. While it is desirable to be as close to the true parameter value as possible, the goal of the procedure is to cut down computation time and to produce close approximations to S-ABC. It would be unreasonable to expect the ABC procedure based on
extrapolated summaries to outperform the S-ABC procedure corresponding to the same summaries in general.
Another notable aspect is the bias of the extrapolated posteriors relative to S-ABC as a function of the accuracy of the extrapolation. This was apparent in the citation network application, where the relative bias changed noticeably with
inclusion/exclusion of triangles as a summary. Improving this accuracy should be the main focus of future work in this area.

Lastly, while not the topic of this paper, the bias of ABC posteriors relative to the true parameter value as a function of the selected summaries is an important issue. In our simulations as well as our application, we only used rather limited sets of summaries. Although the corresponding ABC posteriors show noticeable bias, we mainly sought to convey the proof of concept for an extrapolation-based ABC procedure as well as the interplay between S-ABC and the extrapolated counterparts.
In practice, one should seek to select a more holistic set of summaries to minimize bias.

\bibliographystyle{abbrvnat}  
\bibliography{main.bib}

\section*{Acknowledgements}

This project has been supported by U.S. National Institutes of Health awards R01AI138901, U01HG009088, U54GM088558, R37AI051164, R01AI11233,9, the Swiss National Science Foundation awards CR12I1\_156229 and 105218\_163196, as well as the Swiss Data Science Center (SDSC, Project BISTOM C17-12).


\clearpage

{\LARGE \textbf{Appendix}}

\newcommand{\beginsupplement}{%
        \setcounter{table}{0}
        \renewcommand{\thetable}{S\arabic{table}}%
        \setcounter{figure}{0}
        \renewcommand{\thefigure}{S\arabic{figure}}%
     }
     
\beginsupplement

\appendix

\section{Supplementary Analyses for the Simulation Study}
\label{appendix:sec:A}

\subsection{Additional Covariance Functions for Gaussian Processes}
\label{appendix:subsec:A.1}

To complete the DMC simulation example, here we provide additional Gaussian process (GP) results when using the same mean function but employing different kernels.
Our earlier choice was to use an additive combination of a linear kernel and a radial basis function (RBF) kernel.
In addition to this sum-based version, we also study results using only the linear part, as well as a multiplicative combination of the linear and RBF kernels.
When evaluated at $n_1$ and $n_2$ nodes, for the average degree (denoted $s_1$) and number of triangles (denoted $s_2$), the linear kernel is expressed as follows:
\begin{align*}
    \Cov(s_1(n_1),s_1(n_2)) &= (\alpha \sqrt{n_1} \sqrt{n_2} + \gamma) + \sigma^2 \mathds{1}_{\{n_1 = n_2\}},\\
    \Cov(s_2(n_1), s_2(n_2)) &= (\alpha n_1 n_2 + \gamma) + \sigma^2 \mathds{1}_{\{n_1 = n_2\}},
\end{align*}
while for the linear times RBF kernel:
\begin{align*}
    \Cov(s_1(n_1),s_1(n_2)) &= (\alpha \sqrt{n_1} \sqrt{n_2} + \gamma) \times \exp \left( -\frac{(n_1 - n_2)^2}{2\rho^2} \right) + \sigma^2 \mathds{1}_{\{n_1 = n_2\}},\\
    \Cov(s_2(n_1), s_2(n_2)) &= (\alpha n_1 n_2 + \gamma) \times  \exp \left( -\frac{(n_1 - n_2)^2}{2\rho^2} \right) + \sigma^2 \mathds{1}_{\{n_1 = n_2\}},
\end{align*}
with $\alpha$, $\gamma$, $\rho$, $\sigma^2$ some positive parameters, specific to each kernel, and $\mathds{1}$ being the indicator function. The prior for each hyper-parameter is again a standardized Gaussian distribution with identical truncation as specified in the main text. Moreover, we still consider the two possibilities to deduce the ABC posteriors, either using the densities of the reconstructed normal distributions (without inflated covariance matrices), or based on the distances between GP posterior means and observed summaries.

The linear version is relevant as it provides covariance matrices similar to the ones determined empirically (Main Text, Figure 3), however it might lack flexibility. Regarding the linear times RBF kernel, it has the advantage of providing marginal variances that increase linearly with the number of nodes. It implies that the reconstructed normal distributions are less likely to all give zero density to the observed summary statistics, resulting in posteriors biased toward the prior means.

\begin{figure}
\centerline{
\includegraphics[width=1.02\linewidth, trim={3.7cm 1.8cm 3.5cm 5cm},clip]{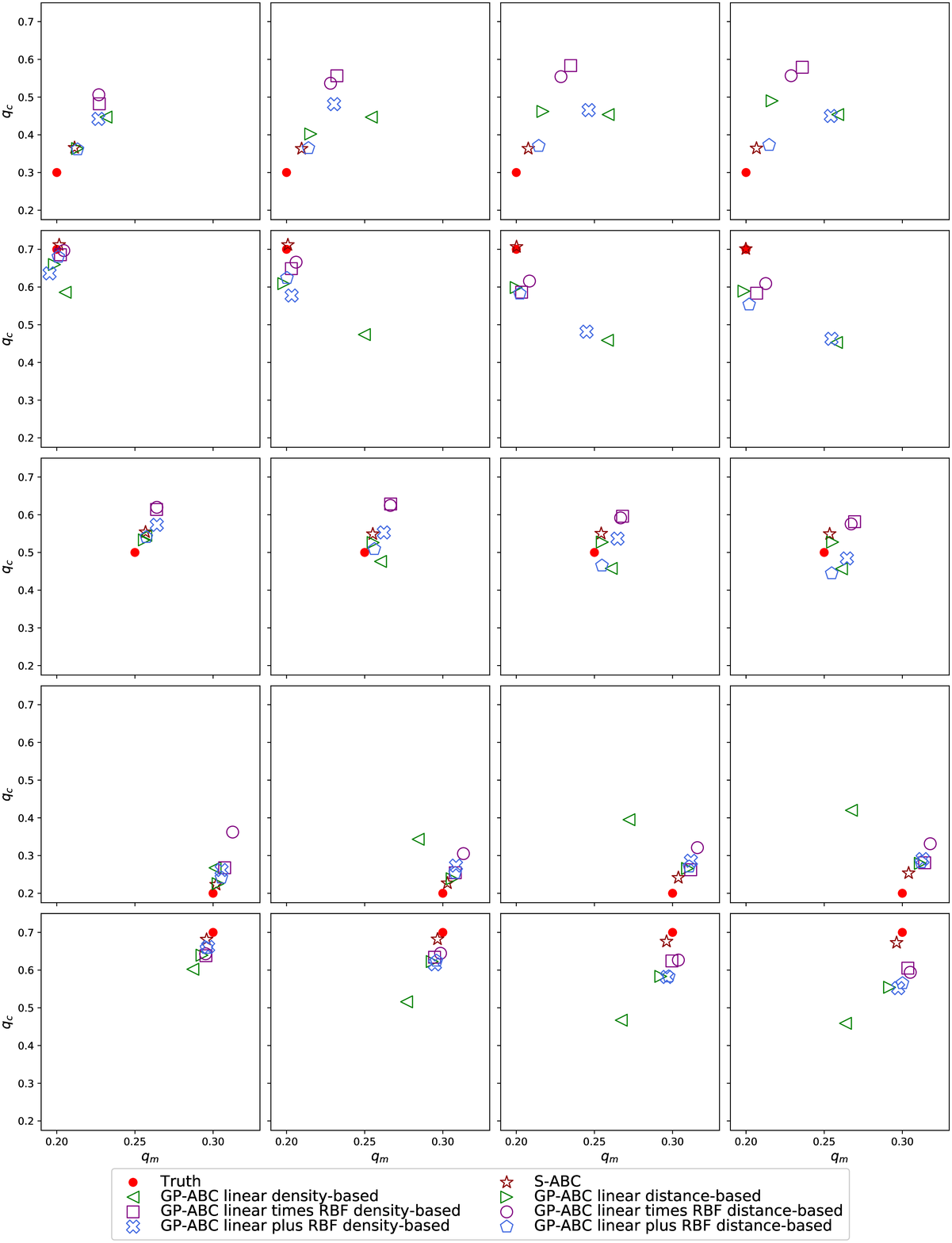}
}
\caption{Average posterior means provided by the standard ABC (S-ABC), and the GP-based methods with the different kernel functions. The red dot denotes the true parameter value. Each row corresponds to different values of the true parameter $(q_{m},q_{c})\in\{(0.2, 0.3)$, $(0.2, 0.7)$, $(0.25, 0.5)$, $(0.3, 0.2)$, $(0.3, 0.7)\}$, each column corresponds (from left to right) to values of $n_o=$ $1000$, $2000$, $5000$, $10000$.}
\label{supfig:GP:post-means}
\end{figure}

\begin{figure}
\centering
\includegraphics[width=\linewidth, trim={3.25cm 2cm 3.5cm 5cm},clip]{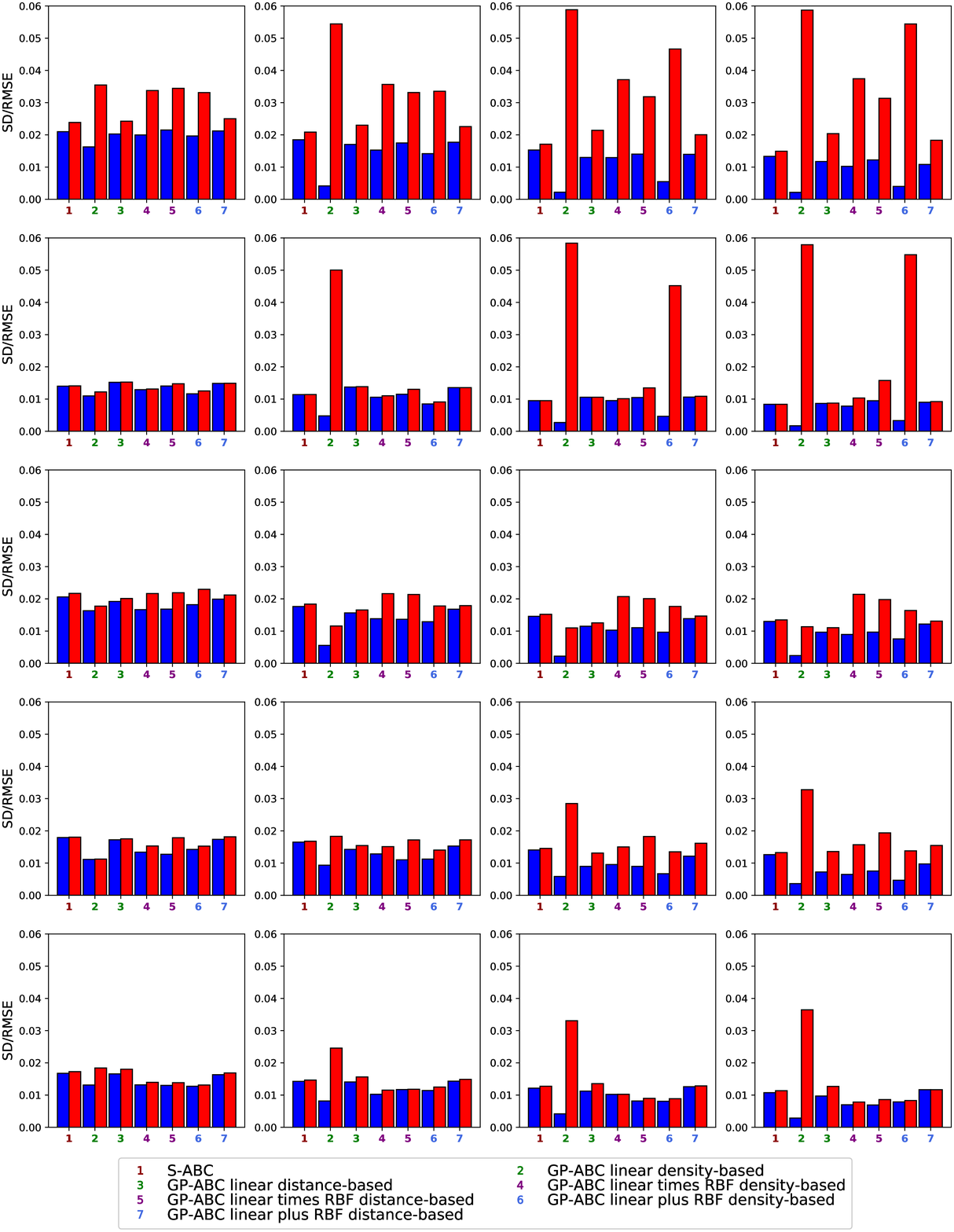}
\caption{Standard deviation (SD, left bar of each pair) and root mean square error (RMSE, right bar of each pair) of the different estimators employing GPs for $q_{m}$.
Each row corresponds to different values of the true parameter $(q_{m},q_{c})\in\{(0.2, 0.3)$, $(0.2, 0.7)$, $(0.25, 0.5)$, $(0.3, 0.2)$, $(0.3, 0.7)\}$, each column corresponds (from left to right) to values of $n_o=$ $1000$, $2000$, $5000$, $10000$.}
\label{supfig:GP:SD-RMSE-qm}
\end{figure}

\begin{figure}
\centering
\includegraphics[width=\linewidth, trim={3.25cm 2cm 3.5cm 5cm},clip]{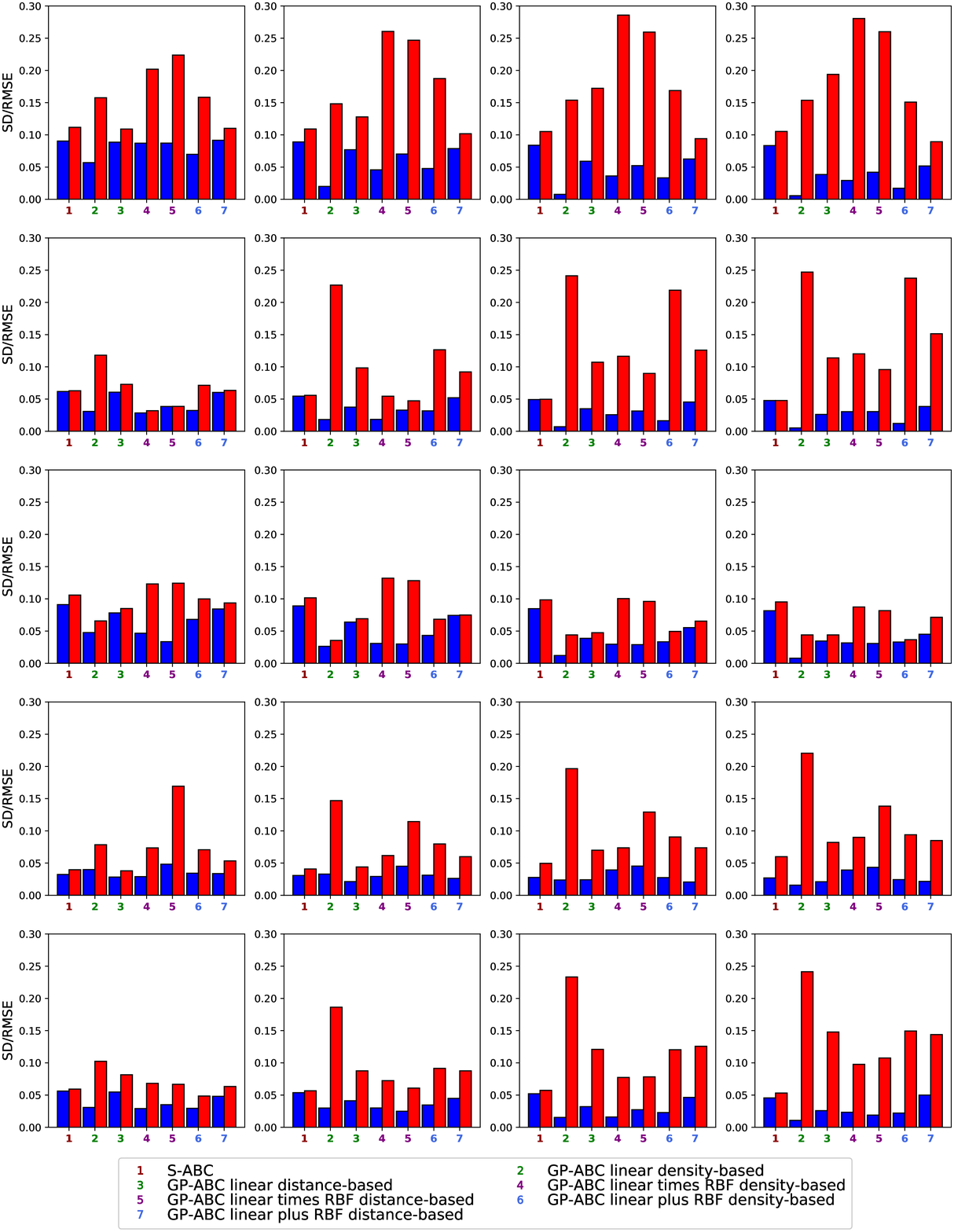}
\caption{Standard deviation (SD, left bar of each pair) and root mean square error (RMSE, right bar of each pair) of the different estimators employing GPs for $q_{c}$.
Each row corresponds to different values of the true parameter $(q_{m},q_{c})\in\{(0.2, 0.3)$, $(0.2, 0.7)$, $(0.25, 0.5)$, $(0.3, 0.2)$, $(0.3, 0.7)\}$, each column corresponds (from left to right) to values of $n_o=$ $1000$, $2000$, $5000$, $10000$.}
\label{supfig:GP:SD-RMSE-qc}
\end{figure}

Figures \ref{supfig:GP:post-means}, \ref{supfig:GP:SD-RMSE-qm} and \ref{supfig:GP:SD-RMSE-qc} show the bias as well as standard deviation (SD) and root mean squared error (RMSE) for the standard ABC (S-ABC) and the GP-based methods with the different kernels.
These figures illustrate that the posterior means, using the linear kernel combined with density-based estimators, are always biased toward the prior means. This is explained by the GP posterior variances  which are here extremely small. The distance-based version performs decently, though it is often outperformed by the linear plus RBF adaptation. The linear times RBF kernel shows great performance for all pair of parameters excepted the first one, for both density-based and distance-based methods, suggesting a large bias of the posterior GP mean. Finally, we can say that there is no specific kernel that outperforms all others for all true parameter pairs. This highlights the importance of selecting a flexible kernel, able to adjust to different shapes of tracked summaries, in different regions of the parameter space.

\subsection{Averaged Sample Triangle Counts}
\label{appendix:subsec:A.2}

Similarly to RE-ABC, which is subsampling 100 nodes to determine sample triangle count, we propose to reduce the variability of this summary by sampling it multiple times (10 or 20), and using the average value as a replacement for the triangle-based summary in the analysis. These are denoted 10 or 20 avg RE-ABC in the following.
Excepted for this change of summary statistic, the extrapolation setting is the same as in the main text.

\begin{figure}
\centerline{
\includegraphics[width=1.02\linewidth, trim={3.7cm 1.8cm 3.5cm 5cm},clip]{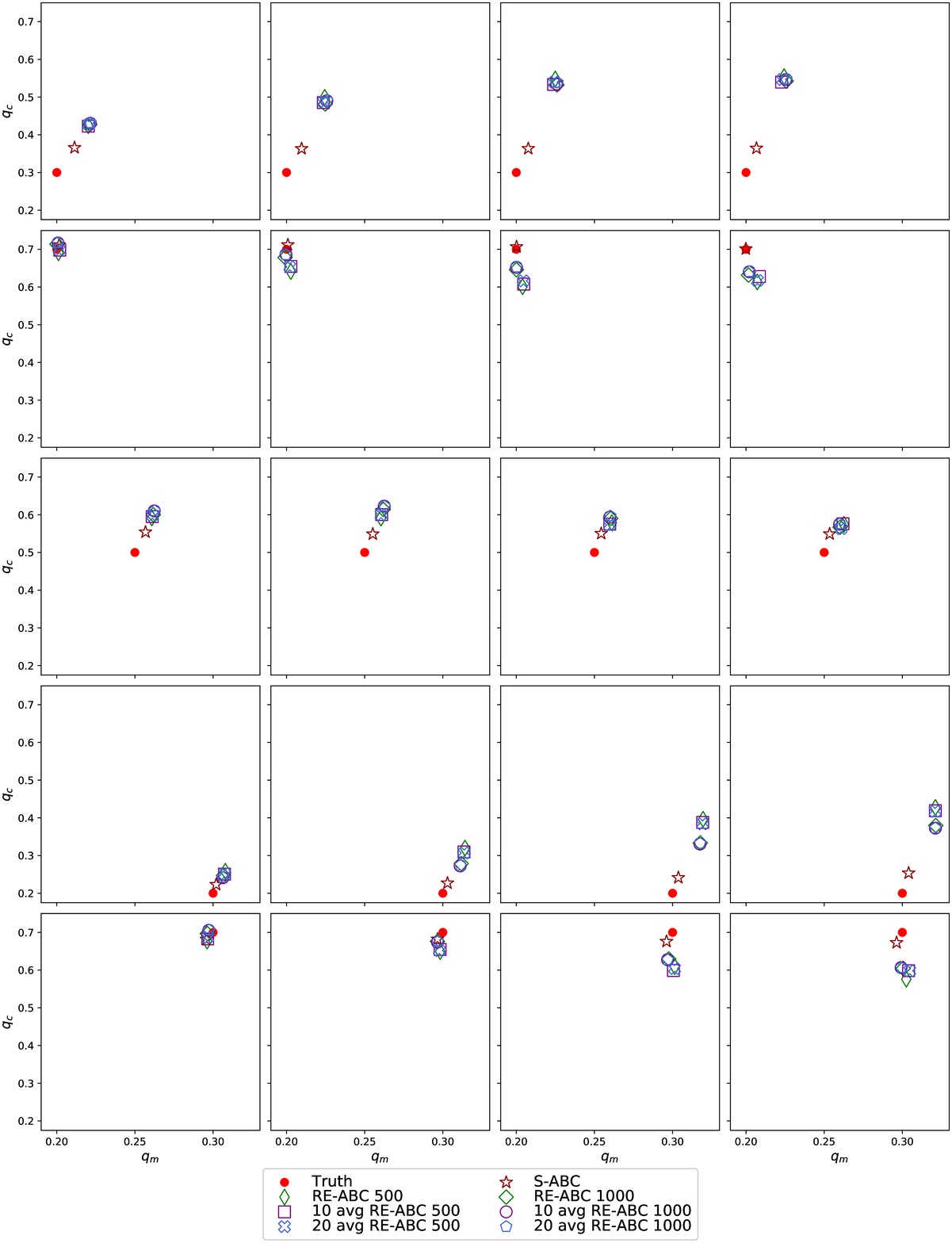}
}
\caption{Average posterior means provided by the standard ABC (S-ABC), and least-squares based methods when employing a single sample triangle count, or the averaged sample triangle count obtained over 10 or 20 replicates. The red dot denotes the true parameter value. Each row corresponds to different values of the true parameter $(q_{m},q_{c})\in\{(0.2, 0.3)$, $(0.2, 0.7)$, $(0.25, 0.5)$, $(0.3, 0.2)$, $(0.3, 0.7)\}$, each column corresponds (from left to right) to values of $n_o=$ $1000$, $2000$, $5000$, $10000$.}
\label{supfig:sample-based:post-means}
\end{figure}

\begin{figure}
\centering
\includegraphics[width=\linewidth, trim={3.25cm 2cm 3.5cm 5cm},clip]{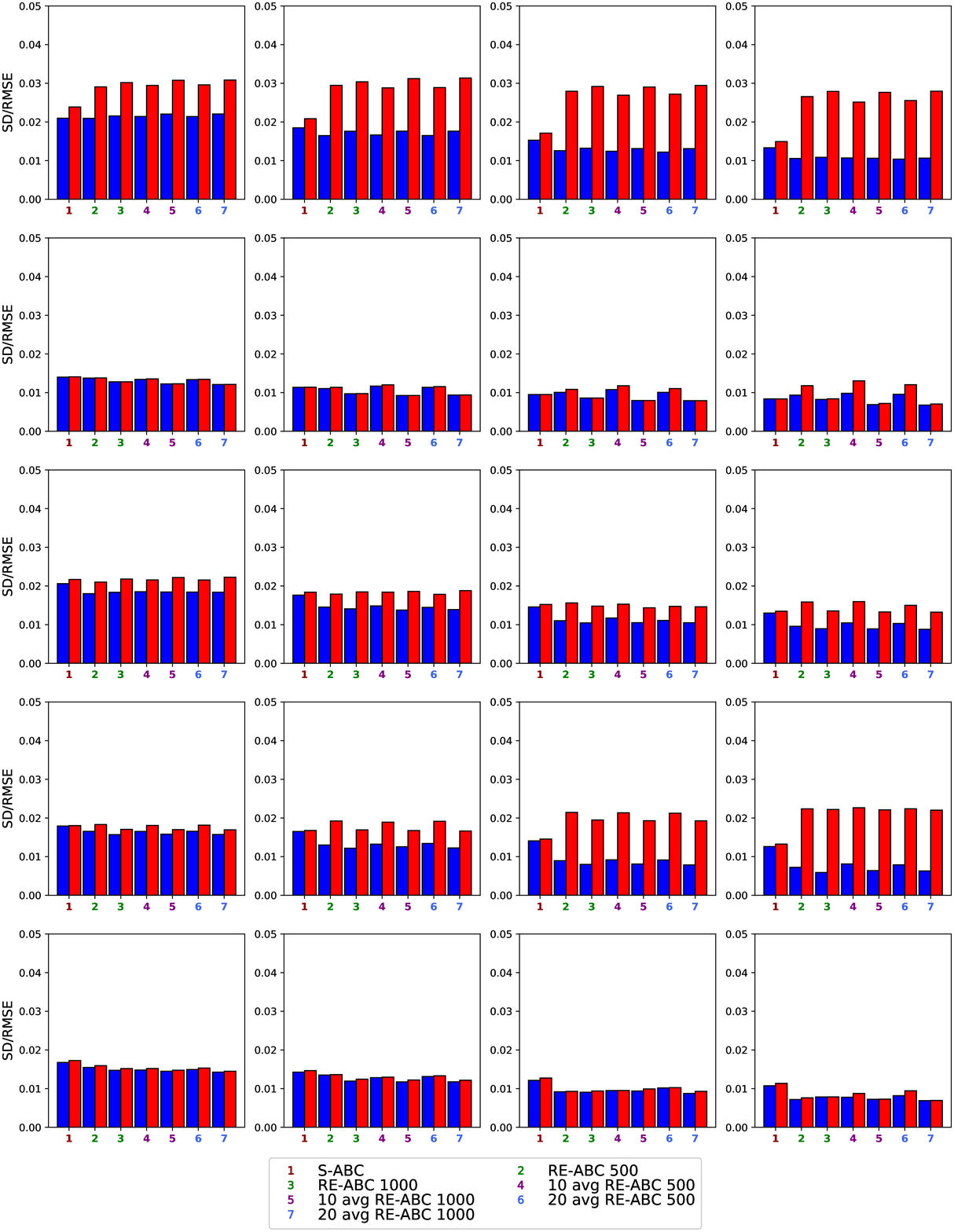}
\caption{Standard deviation (SD, left bar of each pair) and root mean square error (RMSE, right bar of each pair) of the different estimators for $q_{m}$.
Each row corresponds to different values of the true parameter $(q_{m},q_{c})\in\{(0.2, 0.3)$, $(0.2, 0.7)$, $(0.25, 0.5)$, $(0.3, 0.2)$, $(0.3, 0.7)\}$, each column corresponds (from left to right) to values of $n_o=$ $1000$, $2000$, $5000$, $10000$.}
\label{supfig:sample-based:SD-RMSE-qm}
\end{figure}

\begin{figure}
\centering
\includegraphics[width=\linewidth, trim={3.25cm 2cm 3.5cm 5cm},clip]{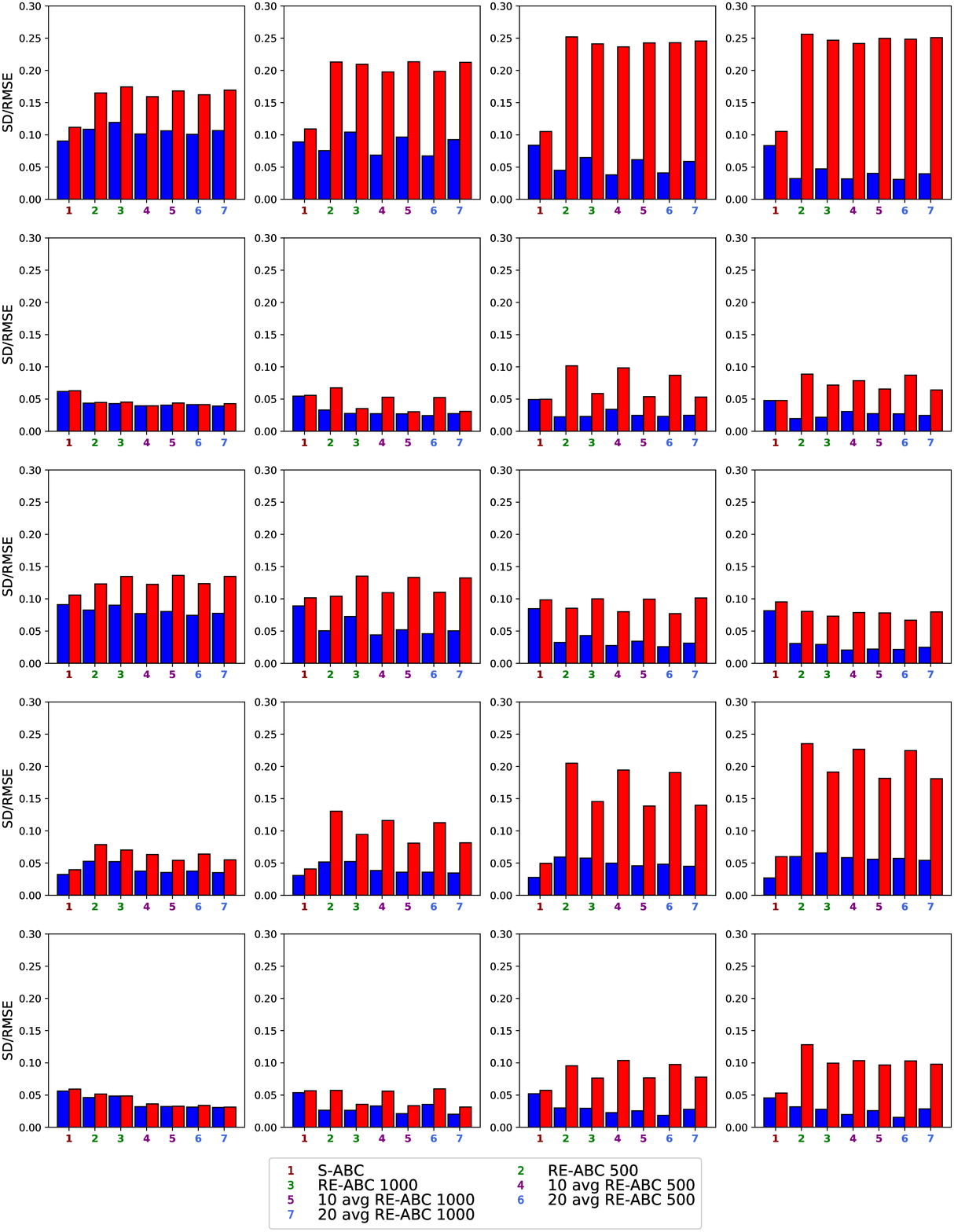}
\caption{Standard deviation (SD, left bar of each pair) and root mean square error (RMSE, right bar of each pair) of the different estimators for $q_{c}$.
Each row corresponds to different values of the true parameter $(q_{m},q_{c})\in\{(0.2, 0.3)$, $(0.2, 0.7)$, $(0.25, 0.5)$, $(0.3, 0.2)$, $(0.3, 0.7)\}$, each column corresponds (from left to right) to values of $n_o=$ $1000$, $2000$, $5000$, $10000$.}
\label{supfig:sample-based:SD-RMSE-qc}
\end{figure}

Figures \ref{supfig:sample-based:post-means}, \ref{supfig:sample-based:SD-RMSE-qm} and \ref{supfig:sample-based:SD-RMSE-qc} show the bias, standard deviations (SD) as well as root mean squared errors (RMSE) for S-ABC and RE-ABC when summaries are tracked up to $n_s=500$ or $n_s=1000$ nodes. These figures highlight little differences in terms of posterior estimates when using either 1, 10 or 20 replicates. Such similarity can be explained by very similar fits between the least-squares adjusted polynomial functions for the sample triangle count, and thus for the extrapolated summary values as well. As expected, we again observe a small gain when tracking up to $n_s=1000$ nodes compared to $500$.

\end{document}